\documentclass[twocolumn,superscriptaddress]{revtex4}
\usepackage{amsmath, amsfonts, amssymb, graphicx, color,amssymb,comment,algorithm,algpseudocode,xcolor,chngcntr}

\DeclareMathOperator*{\argmax}{arg\,max}

\newcommand{\PP}{\color{black}}
\newcommand{\HM}{\color{black}}

\newcommand{\lastequal}{Corresponding authors. These authors contributed equally.}

\begin{document}

 % title
\newcommand{\deftitle}{{Energy-based generative models for monoclonal antibodies}}

\title{\deftitle}

\author{Paul Pereira}
\affiliation{Laboratoire de physique de l'\'Ecole normale sup\'erieure,
  CNRS, PSL University, Sorbonne Universit\'e, and Universit\'e de
  Paris, 75005 Paris, France}\affiliation{Sanofi Vitry-sur-Seine, France}
\author{Herve Minoux}
\affiliation{Sanofi Vitry-sur-Seine, France}
\author{Aleksandra M. Walczak}
\thanks{\lastequal}
\affiliation{Laboratoire de physique de l'\'Ecole normale sup\'erieure,
  CNRS, PSL University, Sorbonne Universit\'e, and Universit\'e de
  Paris, 75005 Paris, France}
\author{Thierry Mora}
\thanks{\lastequal}
\affiliation{Laboratoire de physique de l'\'Ecole normale sup\'erieure,
  CNRS, PSL University, Sorbonne Universit\'e, and Universit\'e de
  Paris, 75005 Paris, France}

\begin{abstract}
 % abstract
Since the approval of the first antibody drug in 1986, a total of 162 antibodies have been approved for a wide range of therapeutic areas, including cancer, autoimmune, infectious, or cardiovascular diseases. Despite advances in biotechnology that accelerated the development of antibody drugs, the drug discovery process for this modality remains lengthy and costly, requiring multiple rounds of optimizations before a drug candidate can progress to preclinical and clinical trials.
This multi-optimization problem involves increasing the affinity of the antibody to the target antigen while refining additional biophysical properties that are essential to drug development such as solubility, thermostability or aggregation propensity. Additionally, antibodies that resemble natural human antibodies are particularly desirable, as they are likely to offer improved profiles in terms of safety, efficacy, and reduced immunogenicity, further supporting their therapeutic potential.
In this article, we explore the use of energy-based generative models to optimize a candidate monoclonal antibody. We identify tradeoffs when optimizing for multiple properties, concentrating on solubility, humanness and affinity and use the generative model we develop to generate candidate antibodies that lie on an optimal Pareto front that satisfies these constraints.

\end{abstract}

\maketitle

\section{Introduction}

 % intro
Monoclonal antibodies are an important type of biological drug, with 162 therapeutic antibodies currently approved used to treat a variety of diseases for different therapeutic areas such as cancer or inflammation. Their development however is costly, time consuming and prone to failure. Antibodies are Y-shaped proteins that specifically bind to an antigen. {\PP The number of possible antibodies that can be designed is too large to search exhaustively when trying to find an antibody that will bind sufficiently strongly to a target antigen.}  
In addition to affinity, there are a number of other desirable properties that the candidate antibody must have in order to function properly as a drug. For example, it is desirable for the antibody not to have exposed hydrophobic patches that could cause aggregation issues and to be soluble. The antigenicity of the candidate antibody must also  be as low as possible in order to avoid an immune reaction in the patient. These constraints further reduce the number of drug candidates and must be taken into account as early as possible in the development process.

We focus on the development of energy-based generative models for monoclonal antibodies. Our work falls within a broad range of methods utilizing wet-lab experiments and machine learning models to optimize antibodies for drug development. In-vitro assays can be used today to characterize antibodies and filter-out ones with poor affinity to the target antigen or poor developability properties \cite{jain2017biophysical}. The development of high-throughput assays has made it possible to characterize large numbers of sequences in batches. Methods like phage or yeast display \cite{mcmahon2018yeast,ledsgaard2018basics} can be used to identify a few antibodies that bind to a target antigen out of a large library of diverse antibodies, or to optimize such candidates by comparing many different mutants of the same wild type sequence \cite{ye2022improving}. However, high throughput methods tend to provide less accurate information than lower-throughput methods such as surface plasmon resonance \cite{hearty2012measuring} and can still take a long time to run compared to in-silico methods.

Machine learning models can been trained on the data generated by these wet-lab experiments~\cite{khetan2022current}. These models can then be used to evaluate the properties of large libraries of antibodies in order to filter-out poor drug candidates and reduce the list of candidates to be tested using wet-lab experiments. This process may lowers the experimental cost, allowing for the use of lower-throughput methods when possible. For example Ref. \cite{mason2021optimization} used affinity information about $1 \times 10^4$ variants of the clinical antibody trastuzumab binding to Her2 to train a machine learning model that was then used to predict the affinity of $1 \times 10^8$ trastuzumab  variants  and to identify top binders. The top predicted binders were further tested using in-silico models for viscosity, clearance, solubility and immunogenicity to end up with a few thousands highly optimized antibodies. While this approach has demonstrated its efficacy, its success relies on optimal candidates being already presents in the library of candidates. This is not guaranteed and depends on the way the library is designed.

Alternatively, generative algorithms directly generate optimized candidate antibodies. 
 Biswas et al~\cite{biswas2021low} successfully combined a predictive model for the fluorescence of green fluorescent proteins and energy based sampling methods to generate new amino sequences of GFP from Aequorea victoria with increased fluorescence. Jain et al~\cite{jain2022biological} demonstrated the ability of autoregressive models to learn to generate amino acid sequences optimized with respect to the output of a proxy predictive model for 3 different tasks (anti-microbial peptide, TFBind 8 and GFP). More recently Bennet et al~\cite{bennett2024atomically} demonstrated the ability of their structure generative method to design de-novo single domain VHH that bind to target antigens for which the epitope is specified.
 
We develop and analyze a method for single round optimization of monoclonal antibodies.
We start from a wild type (WT) sequence of the variable part of an antibody that binds to a target antigen.
While this WT antibody is functional (i.e. binds to the target antigen), we want to improve some of its properties: further increase its affinity to the target antigen or increase its solubility. Furthermore, we  want to improve these properties by performing a small number of mutations, in order to prevent the loss of functionality. Finally, in order to decrease the antigenicity of the antibody, we aim to generate antibodies that are as human-like as possible. In short, we explore the tradeoffs generative models need to consider to produce candidate antibodies that satisfy properties needed in practical antibody discovery: solubility, lack of antigenicity and strong affinity.

\section{Results}

 % background
\subsection{Multiple objective optimization}

Each antibody is made of 2 copies of a heterodimer each composed of a heavy chain and a light chain. The variable regions of the antibodies contain the complementarity determining regions (CDR), flexible loop structures that make up a large part of the paratopes (the section of the antibody that is in contact with the antigen). The heavy chain and light chain each contain 3 CDR regions: CDRH1, CDRH2 and CDRH3 for the heavy chain and CDRL1, CDRL2 and CDRL3 for the light chain. {\PP In this work, we will restrict the optimization to the heavy chain sequence for simplicity although our method can easily be modified to include the light chain.}

We use a multiple objective optimization approach. We consider a set of biophysical properties $f_p(x)$, assumed to be a function of the antibody sequence $x$, which we wish to maximize all at once. We expect trade-offs to emerge since optimizing some properties may come at the cost of others. We look for Pareto optimal solutions, defined as sequences which cannot be improved in any property without being degraded in another one. Combinations of biophysical properties achieved by Pareto-optimal sequences form the Pareto front.

We do not have access to the functions $f_p$ directly for arbitrary sequences $x$. Instead, we will exploit predictive machine learning models $\hat f_p$ trained on large datasets to generate sequences that are close to the Pareto front. 
We are interested in the single-round setting: we assume that once our generative method has selected candidates for us to test, we will be able to validate those candidates using wet-lab experiments only once, with a fixed budget of $B$ sequences to be tested.
Since we do not expect the models $\hat f_p$ to perfectly reproduce the desired properties, we wish to generate a diverse set of $B$ candidates to maximize the probability that at least one them will pass validation. We further require that these sequences be different from the ones included in the datasets used to train the models.

In the following, we will focus on two main biophysical properties: solubility (sol) and affinity (aff).
Our generative process incorporates the humanness of sequences through the distribution of natural antibodies $p_{\rm HUM}(x)$, as given by an auto-regressive transformer previously trained on 558 million human heavy chain sequences from the OAS database \cite{shuai2021generative} \cite{olsen2022observed}. We define $p(x)$ for our generative model to be as close as possible to $p_{\rm HUM}$ (as measured by the Kullback-Leibler divergence) while maximizing the mean values of $f_{\rm sol}$ and $f_{\rm aff}$. This gives us the explicit expression:
\begin{align}
  p(x)&=\frac{1}{Z}p_{\rm HUM}(x)e^{-E(x)/T}\label{eq:bolz}\\
  E(x)&=-w \hat f_{\rm aff}(x) -(1-w) \hat f_{\rm sol}(x),
\end{align}
where $Z$ is a normalization, $T$ may be interpreted as a temperature controling to cloneness to the Pareto front, and $0\leq w\leq 1$ is a weight controling the importance of affinity versus solubility in the optimization.

\begin{figure*}
\begin{center}
\includegraphics[width=\linewidth]{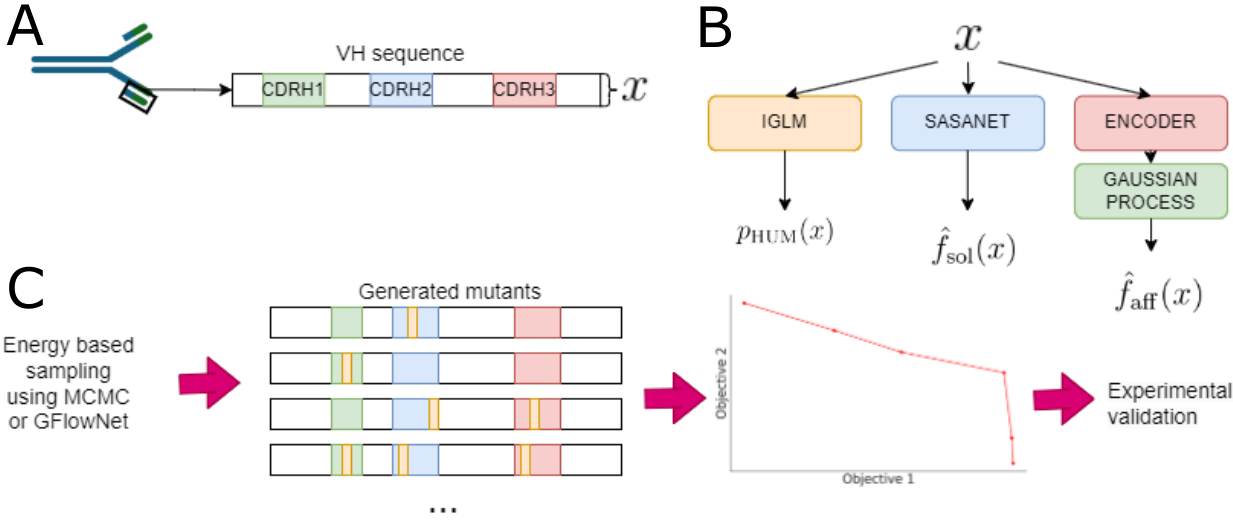}
\caption{Overview of the generative process. A. Our generative model generates the heavy chain sequence of an antibody undergoing lead optimization. The goal is to find a small number of mutations in the CDRs to improve affinity and solubility. B. For humaness, our predictive model is IGLM, an autoregressive transformer which outputs a log probability estimate of the input heavy chain being human. For solubility we use a convolutional neural network that estimates the per-residue SASA and combines these estimates with hydrophobicity weights. For the affinity we use a Gaussian Process in combination with a protein language model encoder. C Energy based sampling is used to generate mutants of the wild-type sequence with up to $d_lim$ mutations. The closest sequences to the pareto front are selected for wet-lab validation}
\label{fig:outline}  
\end{center}
\end{figure*}

We then use two energy-based generative methods to sample new heavy chain sequences from $p(x)$ (Figure \ref{fig:outline}): Metropolis Hastings (MCMC) and the amortized Monte Carlo method GFlowNet \cite{haarnoja2017reinforcement, bengio2023gflownet}.  GFlowNet has been shown to generate a better and more diverse set of amino acid sequences on  multiple tasks such as generating new anti-microbial peptides \cite{jain2022biological}. Details of these procedures are given in the Methods section.

 % fig0

Below we apply this approach to generate optimized binders to a HR2 region of the SARS-CoV-2 spike protein peptide (CB-119) whose amino acid sequence is PDVDLGDISGINAS.

 % model

\subsection{Affinity model}

Antibody-antigen affinity is often defined in terms of the dissociation constant ($K_d$) between an antibody and its antigen. High affinity means low $K_d$, so we define $f_{\rm aff}=\log (1\,{\rm nM}/K_d)$.
In order to train our affinity prediction model, we use the dataset from Engelhart and al. \cite{engelhart2022dataset}. A library of variants was built by first identifying a pair of heavy and light chain that produce an antibody (Ab-14) that binds to CB-119. {\PP{The heavy chain of Ab-14 has 33 CDR amino acids that were modified to generate 22000 mutants}}:
594 single mutants were generated by doing a saturated deep mutational scan over the CDRH1, CDRH2 and CDRH3.
In addition 3,671 double mutants and 22,188 triple mutants were generated by performing random substitutions in the CDRs. 
The affinity of each sequence was measured 3 times using the AlphaSeq assay to compute their $K_d$ with respect to the CB-119 SARS-CoV-2 peptide. Of the 26453 sequences, only 13921 sequences had affinity measurements. We approximate their true $K_d$ as the average over replicates and removed the sequences for which there are no measurements.

For our affinity prediction method, we use a Gaussian Process. The Gaussian process assumes that outputs (affinities) are drawn from a multivariate Gaussian distribution whose covariance between two outputs depends on the distance between their input sequences $x$, and uses Gaussian integration rules to predict the posterior for the affinity of any sequence $x$ as a function of the training data:
\begin{equation}
f_{\rm aff}(x) \sim \mathcal{N}(\mu_{\rm aff}(x;D),\sigma_{\rm aff}^2(x;D)),
\end{equation}
where $D=(x_i,f_{\rm aff}(x_i))_i$ is the training dataset.

We define multiple models using different embeddings for the amino acid sequences and used a Radial Basis Function (RBF) as our kernel function (see Methods).

The parameters of the kernel function ($\delta$ and $\lambda$) are fitted by minimizing the marginal likelihood loss function (see Eq. \ref{eq:RBF}) for 600 steps using a learning rate of $10^{-3}$ (see 
Methods for details). We utilized the gpytorch \cite{gardner2018gpytorch} framework to implement the Gaussian Process and fit the parameters.

To score sequences, we define 4 different versions of the affinity function to be used in the multiple objective optimization: $\hat f_{\rm aff}(x;\beta)= \mu(x) + \beta \sigma(x)$, with $\beta=-1,0,1,2$.
When $\beta$ is positive we get back the Upper Confidence Bound (UCB) acquisition function \cite{auer2002finite} which gives priority to sequences with high uncertainty and encourages exploration of the sequence space. When $\beta$ is negative, we obtain the Lower Confidence Bound (LCB) acquisition function, a pessimistic estimate of the affinity of the sequence. Using LCB encourages the model to generate sequences close to the ones in the training set 

{\PP We split the dataset into a training and a validation set. We refer to the training set as the set of sequences that are used to fit the parameters of the kernel function and to make a prediction. In order to test the ability of the Gaussian process to correctly predict the affinity to CB-119, we first built a training set of 80\% of D and validate the model on the remaining 20\%}. The Pearson correlation score between {\PP {the log of the measured association constant}} $f_{\rm aff}$ and the predicted affinity $\mu_{\rm aff}$ on the validation set for the best choice of embedding and when training on $80\%$ of the data is $0.58$ (Figure \ref{fig:1}A) showing that the model is able to discriminate between high and low affinity sequences.

Using the protein language model (PLM) ESM2 \cite{lin2023evolutionary} for the embeddings gives a modest improvement over the simpler one-hot vector encoding of the sequences (Fig.~\ref{fig:1} B). We observe no differences between using a smaller version of ESM (ESM-T6) versus a larger network (ESM-T30). Using an antibody specific protein language model (Antiberty) \cite{ruffolo2021deciphering} gives worse results than a more general PLM (Fig.~\ref{fig:1}). We asked whether reducing the amount of training data significantly decreases the performance of the model.
When we reduce the size of the training set to only 30\% of the data, the Pearson correlation coefficient decreases by only $0.02$ when using the ESM-T6 PLM to compute the embedding. Since the running time for the prediction of a Gaussian Process is quadratic in the number of sequences in the training set, evaluating new sequences during the generative processes we use the ESM-T6 embeddings and train the Gaussian Process on $30\%$ of the dataset to limit the computational cost.

\subsection{Solubility model}

The solubility of a monoclonal antibody is related to a different factors that can impact the efficacy of the drug. Antibodies can aggregate when stored in an aqueous solution which will result in a painful reaction when the drug is administered to the patient. In addition, aggregates can become a target for the immune system and increase the antigenicity of the drug. Hydrophobic interaction chromatography (HIC) estimates the solubility of antibodies by measuring the time time it takes the antibody to cross a buffer. Stronger hydrophobic amino acids result in longer times.

We use a method developed by Jain and al.~\cite{jain2017prediction} to estimate the solubility of an antibody based on its sequence. Given a sequence of amino acid of length L: $x = x_1,x_2,...,x_L$, the solubility score of the sequence is defined as

\begin{equation}
    f_{\rm sol}(x) = -\sum_{j = 1}^L {\rm SASA}(j,x_j) {\rm HW}(x_j)+\textrm{const},
\end{equation}
where ${\rm SASA}(i,S_i)$ is the residue solvent accessible surface area of amino acid $a_i$ at position $i$, and ${\rm HW}(x_i)$ is one of 20 hydrophobic weights describing the hydrophobicity of the amino acid $x_i$. We use the solubility weights provided in Jain et al~\cite{jain2017prediction} and train a convolutional neural network to predict the per-residue SASA score of each amino acid in an antibody heavy chain sequence (see Section Methods B). 

Since the per-residue SASA score is not available from the sequence alone, Jain and al. compute the per residue SASA for 902 antibody structures identified from the RCSB using the Shrake-Rupley algorithm~\cite{shrake1973environment} and train a random forest regressor to predict the per residue SASA of residues in the variable region from the sequence alone. They then use a private dataset of 5000 antibody sequences for which the HIC retention time (RT) was measured to learn the $HW$ weights using logistic regression. We re-implemented their method but replaced the random forest regressor with a deep convolutional neural network  NanoNet \cite{cohen2022nanonet}, a structure prediction method for antibodies, and included more structures taken from the SAbDab database \cite{dunbar2014sabdab}. Since this architecture is able to accurately predict the structure of antibodies, we hypothesized it may predict the per-residue SASA score. We modified the last layer of the network  to output a single value corresponding to the SASA score for each residue instead of the 15 values corresponding to the coordinates of the backbone residues (see section \ref{sec:solpred}). 

\begin{figure*}
\begin{center}
\includegraphics[width=\linewidth]{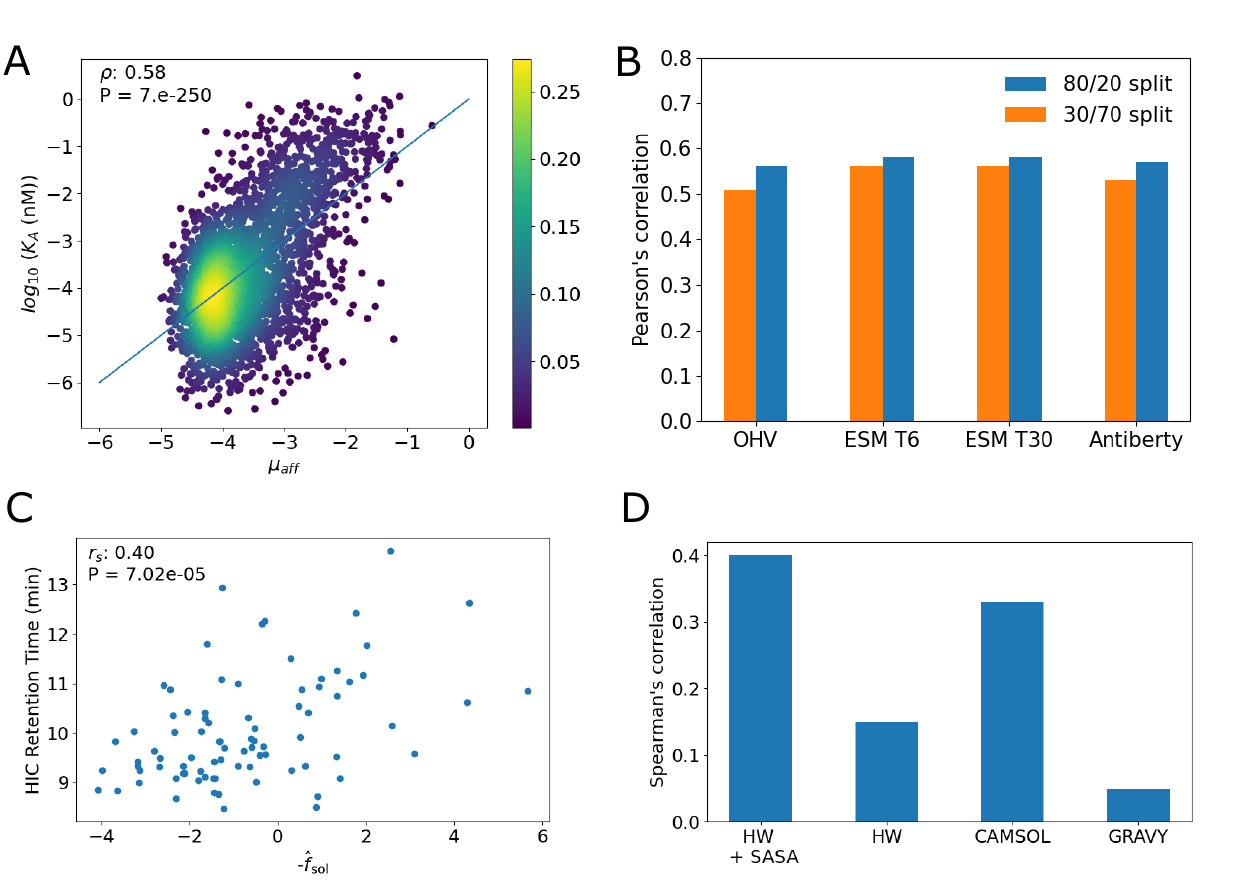}
\caption{ A. Density plot showing results from predicting KD using Gaussian Process B. Pearson's correlation coefficient for different choices of embedding C. Scatter plot comparing HIC RT to Solubility score for 83 monoclonal antibodies D. Pearson's correlation score for different solubility methods. HW + SASA refers to the method we retain for the generative process, HW is similar to HW + SASA except the predicted SASA for each residue is set to 1. Camsol refers to the Camsol method \cite{sormanni2015camsol} and Gravy refers to the biopython hydrophobicity score function.}
\label{fig:1}  
\end{center}
\end{figure*}

We use a dataset containing 137 clinical stage antibodies for which the HIC RT was computed ~\cite{jain2017biophysical}. We used the Thera-SAbDab database~\cite{dunbar2014sabdab} to identify the antibodies in this dataset whose structures were used to train the SASA prediction model and excluded them, leaving us with 83 clinical stage antibodies. The Spearman correlation coefficient between the HIC RT and the predicted solubility score on the 83 clinical stage antibodies is 0.40. (Fig.~\ref{fig:1} C). We compare the performance of our solubility model to other models used to evaluate the solubility of antibodies (Fig.~\ref{fig:4} D). Removing the SASA in the predictive models leads to a drop in the Pearson correlation score to $0.3$. The SASA model we use is comparable in performance to the commonly used CamSol~\cite{sormanni2015camsol} \cite{sormanni2017rapid}, a sequence based model that returns a hydrophobicity score for every amino acid in the sequence that can be averaged to produce an overall score. Augmenting Camsol with our SASA prediction model only provides a marginal improvement ($0.4$ vs $0.38$). Finally, our method performs much better than the Gravy biopython score~ \cite{khan2022antbo} ($0.39$ vs $0.18$). 

 % fig1

 % results

\subsection{Pareto optimal binders to CB-119 peptide}

To generate a diverse set of antibodies binding the CB-119 peptide and optimized for solubility, we generate 20 sets of sequences with 5 different weights $w =0.85,0.875,0.9,0.95,1.0$ and the 4 choices of $\beta=-1,0,1,2$. We restrict ourselves to sequences at most 6 mutations away from the wildtype (WT) AB-14 (see Discussion).

We define the distance to the Pareto front as:
\begin{equation}
    d_{\rm P}(x) = \min_{x' \in {\rm PO}} \frac{(f_{\rm aff}(x) - f_{\rm aff}(x'))^2}{\sigma_{\rm aff}^2} + \frac{(f_{\rm sol}(x) - f_{\rm sol}(x')^2}{\sigma_{\rm sol}^2},
\end{equation}
where $\sigma^2_{\rm aff}$ and $\sigma^2_{\rm sol}$ are the variances of the affinities and solubilities of the generated sequences.
Top sequences are defined as the $B$ sequences with smallest $d_{\rm P}$, where $B$ is the number of sequences we wish to generate and generally corresponds to a budget of sequences that can be experimentally validated following the generation process.

 % fig2
\begin{figure*}
  \includegraphics[width=\linewidth]{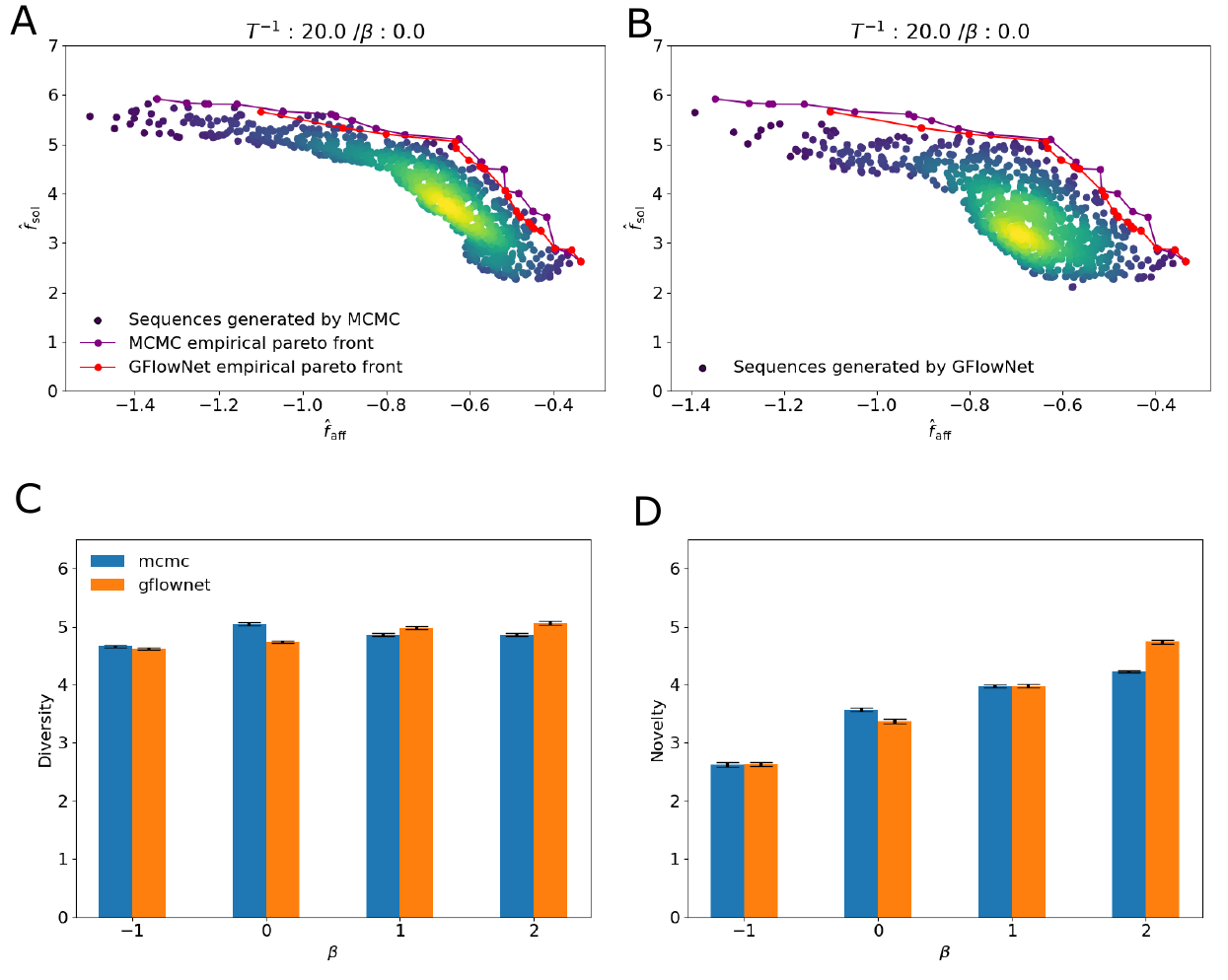}
\caption{A. Density plots of 1000 sequences sub-sampled from the set generated by Metropolis-Hastings at inverse temperature $T^{-1} = 20$ and $\beta = 0.0$. B. Density plots of 1000 sequences sub-sampled from the set generated by the GFlowNet at inverse temperature $T^{-1} = 20$ and $\beta = 0.0$. C. Comparison of diversity based on generative method and choice of $\beta$. D. Comparison of novelty based on generative method and choice of $\beta$}.
\label{fig:2}
\end{figure*}

Fig.~\ref{fig:2} A and B show the empirical Pareto fronts from the generative process for an inverse temperature of $T^{-1}=20$  and a $\beta = 0$. The Pareto fronts for $\beta = 2.0$, $\beta$ = 1 and $\beta$ = -1 are included in SI Fig.~\ref{figsi4}. The Pareto fronts generated by Metropolis-Hastings sampling and the GflowNet are similar for $\beta = 0$ and $\beta=-1$, suggesting neither method is better than the other one at finding optimal sequences. For $\beta = 1$ and $\beta = 2$,  Metropolis-Hastings sampling generates sequences with higher affinity and solubility than the GFlowNet. 
Additional training of the GFlowNet could possibly improve its performance, {\PP {for example by doubling the number of training steps from 8000 to 16000.}} For $\beta = 0$, the  generated sequences that have the highest affinity have a solubility score of around $2.5$ while the generated sequences that have the highest solubility have an affinity of $10^{-1.4}\approx 0.04$ nM. {\PP As a point of comparison, the 83 therapeutic antibodies on which we evaluated our solubility prediction method have an average solubility score $\hat f_{\rm sol}$ of $\approx 0.7$}.
The shape of the Pareto fronts show a trade off between the solubility and the affinity score. The Pareto front obtained using the predictive models suggests that increasing the solubility score from $2$ to $6$, a solubility score higher than any of the therapeutic antibodies used to validate the model, decreases the affinity by one order of magnitude. 

To evaluate the novelty and diversity of a set of generated sequences $D_{\rm gen}$, we follow Jain et al~\cite{jain2022biological} and define a diversity index given by the mean Hamming distance within the generated dataset:
\begin{equation}
  {\rm Diversity}(D_{\rm gen}) = \frac{1}{N_{\rm gen}(N_{\rm gen}-1)}\sum_{(x,x') \in D_{\rm gen}} d_H(x,x'),
\end{equation}
where $d_H(x,x')$ is the Hamming distance between $x$ and $x'$, and $N_{\rm gen}=|D_{\rm gen}|$.
Novelty is defined as the mean distance to the original dataset:
\begin{equation}
    {\rm Novelty}(D_{\rm gen}) = \frac{1}{N_{\rm gen}}\sum_{x \in D_{\rm gen}} \min_{x'\in D_{\rm aff}}d(x,x'),
\end{equation}
where $D_{\rm aff}$ is the dataset of sequences for which affinity was measured.
Figure \ref{fig:2} C and D shows the diversity and novelty of the sequences generated for both methods and the three choices of $\beta$. 

The average mean pairwise Hamming distance of generated sequences is between $4.5$ and $5$ depending on the choice of $\beta$ and sampling method (Figure \ref{fig:2}~C), compared to the maximum possible value of $12$ {\PP since the exploration is limited to the space of sequences that are at most 6 amino acid away from the wildtype sequence}. Novelty shows an expected dependence on the exploration parameter $\beta$: it ranges from $\sim2.5$ for a pessimistic acquisition function $\beta = -1$ to $\sim 4$ for an exploratory acquisition function $\beta = 1$. Both MCMC and GFlownet sampling methods  generate a diverse and novel set of sequences. In addition, the continuity of the cloud of points shows that the methods can generate points all along the Pareto front from which we can pick the sequences with the desirable trade-off between affinity and solubility.

The inverse temperature parameter $T^{-1}$ provides a balance between diversity and optimality. A low value will lead to more diversity in the generated set but may prevent the method from generating sequences in or near the Pareto optimal set. To verify that $T^{-1}$ is sufficiently high to generate Pareto optimal sequences, we generated $30$ new sets of sequences with $T^{-1}=25$ and $T^{-1}=30$ using MCMC sampling for each choice of $\beta$ and $w$. We then combined all sequences generated for different choices of $w$ into a single set and identified the Pareto optimal set shown in SI~Fig~\ref{figsi1}. Decreasing the temperature does not generate more optimal sequences, suggesting that an inverse temperature of 20 is sufficient to generate the Pareto optimal set.

We evaluated the sequences generated using two other developability scores reported in previous work to evaluate generative methods~\cite{khan2022antbo,li2023machine}: the average charge and instability. For the instability measure, we  use the {\PP{instability\_index()}} Biopython package function, which takes as input an amino acid sequence and returns a real value instability score~\cite{guruprasad1990correlation}. A protein is considered unstable if the instability score is less than $-40$. To estimate the charge of the antibody, we sum the contributions from all amino acids in the heavy chain sequence, adding $+1$ for positively charged amino acids R and K, $+0.1$ for all H and $-1$ for negatively charged amino acids D and E. A good candidate antibody should have a  charge score to be between $-2$ and $2$. The results for $T^{-1} = 20.0$ and all choices of $\beta$ are shown in supplementary information figure \ref{figsi3}. {\PP {The sequences generated by our method have an average charge score within the desired range, with $1.25$ for $\beta = -1$ and $0.5$ for $\beta = 1$. The standard deviation is approximately 1 for all choices of $\beta$ and we observe that a small fraction of the sequences have a charge score higher than the limit of 2. This is likely due to the fact that positively charged amino acid are more hydrophilic and the generative model can generate more soluble antibodies by adding positively charged amino acids to the antibody sequence. In addition, we observe that for all choices of $\beta$, the instability score remains on average slightly below $30$ with a standard deviation of $3$, with the vast majority of sequences having an instability score below the threshold of $40$. In conclusion, these results suggests that our method is able to generate functional proteins, even when evaluated on methods that were not included in the multi-objective optimization function. }}

\subsection{Validation on synthetic dataset}

We have demonstrated the ability of energy based generative methods to generate Pareto optimal sequences with respect to the proxy functions $\hat f_{\rm aff}$ and $\hat f_{\rm sol}$. However, we are not able to experimentally evaluate the $K_d$ of HIC of the sequences to verify that the sequences generated are actually functional.
To circumvent this limitation and test the validity of the method, we designed a synthetic dataset $D_{\rm syn}$ on which to test our approach.

We define an epistatic affinity model as:
\begin{multline}\label{eqn:epistasis}
    f_{\rm aff}(x) = f_{\rm aff}(x^{\rm WT}) + \sum_{i = 1} ^L  h_i(x_i) + \sum_{i<j} J_{ij}(x_i,x_j),
\end{multline}
where $x^{\rm WT}$ is a wild type sequence, $h_i(i)$ is the change in $\log (1/K_d)$ after mutating the amino acid at position $i$ to $x_i$ away from its wildtype value, and $h_i(x_i) + h_j(x_j) + J_{ij}(x_i,x_j)$ is the change caused by mutating two amino acids at position $i$ and $j$ to $x_i$ and $x_j$ away from the wildtype. This epistasis model was previously used to estimate the affinity of antibodies to fluorescein based on a dataset similar in composition to the CB-119 peptide dataset studied above~\cite{adams2019epistasis} and fits within the broader class of epistatic model decomposed by order of interaction, here truncated at second order---check Phillips et al. \cite{10.7554/eLife.71393} and Ranganathan review \cite{poelwijk2016context} on epistasis.

Using this framework, we define two synthetic affinity models in which $h_i(x_i)$ and $J_{ij}(x_i,x_j)$ are sampled from a Gaussian distribution. The first model, referred to as the simple epistasis model has $H(i,S_i) = N(-0.5,0.5)$ and  $J(i,j,S_i,S_j) = N(0.0,0.5)$. The second "hard" model has   $H(i,S_i) = N(0.0,0.5) $ and $J(i,j,S_i,S_j)=N(-0.5,0.5)$. {\PP The probability that a random mutation will be deleterious is higher in the hard case than in the simple case due to the larger impact of the J terms compared to the H terms, which motivates the hard/simple terminology.}

To generate our synthetic dataset, we use the same wildtype sequences as in the SARS-CoV-2 affinity dataset, and compute the affinities (Eq \ref{eqn:epistasis}) of 14660 random variants containing all 660 single mutants, 2100 double mutants and 11900 triple mutants, comparable to the affinity Engelhart dataset~\cite{engelhart2022dataset}. {\PP Each mutant was generated by starting from the wildtype sequence AB-14, randomly selecting a number of positions that we want to mutate and performing a random substitution at each of these positions.} We further add some random noise $\sim N(0,1)$ to each measurement to mimic experimental error. We use this data to train a Gaussian Process as the new predictive affinity model $\hat f_{\rm aff}$, and generate sequences optimized for that objective as well as $\hat f_{\rm sol}$.
For each of the two tasks, we set the inverse temperature $T^{-1}=10$ and generate $6$ sets of sequences with $w \in \{0.85,1.0\}$ and $\beta \in \{-1.0,0.0,1.0,2.0\}$. {\PP We initially tried an inverse temperature of $T^{-1}=20$, however we found that value to be too high and the process  failed to generate diverse sequences. We lowered the inverse temperature to 10 to more closely match the behavior we observed when generating binders to CB-119.}

 % fig3
\begin{figure*}
\begin{center}
\includegraphics[width=\linewidth]{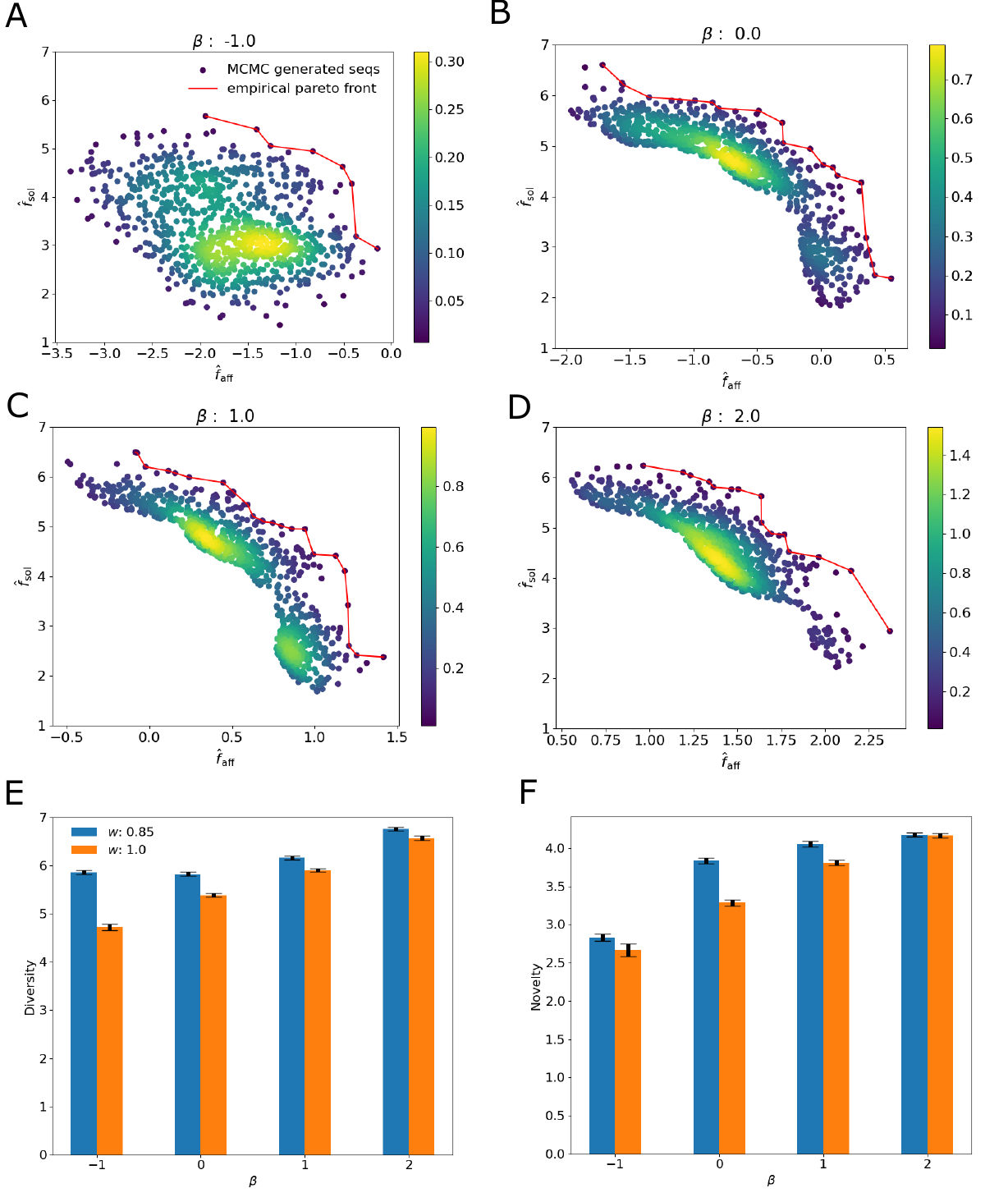}
\end{center}
\caption{A. Density plots of 1000 sequences sub-sampled from the set generated by Metropolis-Hastings at inverse temperature $T^{-1} = 10$ and $\beta = -1.0$ on the simple synthetic task. B. Density plots of 1000 sequences sub-sampled from the set generated by Metropolis-Hastings at inverse temperature $T^{-1} = 10$ and $\beta = 0.0$ on the simple synthetic task. C. Density plots of 1000 sequences sub-sampled from the set generated by Metropolis-Hastings at inverse temperature $T^{-1} = 10$ and $\beta = 1.0$ on the simple synthetic task. D. Density plots of 1000 sequences sub-sampled from the set generated by Metropolis-Hastings at inverse temperature $T^{-1} = 10$ and $\beta = 2.0$ on the simple synthetic task. E,F. Comparison of diversity and novelty for different choices of $\beta$ and $w$.}
\label{fig:3}
\end{figure*}

Figures~\ref{fig:3}~A, B and C show the $1,000$ sequences closest to the empirical Pareto front for the simple task, using MCMC sampling to generate new sequences for the 3 choices of $\beta$. {\PP We observe that for all choices of $\beta$, the method is able to generate a diverse set of sequences along the empirical Pareto front. We observe a clear trade-off between the solubility score and the affinity score, especially for $\beta$ = 0, 1 and 2. However, similarly as for the CB-119 task,  for $\beta = -1$ the method has more difficulty generating many sequences very close to the empirical Pareto front. This result stems from the fact that $\beta = -1$  limits the search space to sequences similar to the ones in $D_{\rm syn}$, decreasing the probability of the generative method to find the most optimal sequences according to $\hat f_{\rm aff}$ and $\hat f_{\rm sol}$}. Furthermore, we observe that the generated sequences are slightly more diverse and novel when $w = 0.85$, for all choices of $\beta$ (Figure \ref{fig:3}~D). The results for the hard epistasis model are shown in supplementary figure \ref{figsi6}. {\PP The results for the hard task are similar to the simple task, except for $\beta = -1.0$ for which the methods fails to generate many diverse solutions. Instead we generate only 17 unique sequences. One possible reason, is the small number of high affinity sequences in $D_{\rm syn}$ with 2 or 3 mutations (since the hard task penalizes the J epistasis terms) (Figure \ref{figsi7} C). This in turn may make it difficult for the Gaussian process to predict with high certainty that sequences with more mutations have a high affinity. The generative process then produces many single mutants which are subsequently removed from our results since all single mutants are already included in $D_{\rm syn}$. From this result, we learn that on tasks where improving the affinity is difficult, i.e. there are not many beneficial mutations, it is especially important to use an optimistic approach to find novel candidates to test.}

In order to investigate the usefulness of our method for antibody lead optimization, we consider the task of generating antibodies with predicted solubility score $\hat f_{\rm sol}$ greater than a threshold $f_{\rm sol}^{\rm min}$ and with a synthetic affinity $f_{\rm aff}(x)$ greater than a threshold $f_{\rm aff}^{\rm min}$.
For each choice of $\beta$ and $w$, we take the set of generated sequences using MCMC sampling and select {\PP the} ones with $\hat f_{\rm sol}> f_{\rm sol}^{\rm min}$. We  rank the selected sequences by $\hat f_{\rm aff}$ and keep the top $B$ sequences, where $B$ is the budget of sequences that can be validated with wet lab experiments. For every threshold $f_{\rm aff}^{\rm min}$ we count the number of sequences, out of the $B$ sequences generated and selected, with synthetic affinity $\hat f_{\rm aff}> f_{\rm aff}^{\rm min}$. We choose thresholds $f_{\rm sol}^{\rm min} = - \infty$ (no solubility threshold) and $f_{\rm sol}^{\rm min}= 4.0$ to explore the differences when optimizing for both affinity and solubility vs simply affinity, and a budget of $B = 500$. $f_{\rm sol}^{\rm min} = 4.0$ is sufficiently high to demonstrate a difference not having a solubility threshold while still generating valid sequences. 

 % fig4
\begin{figure*}
\begin{center}
\includegraphics[width=0.8\linewidth]{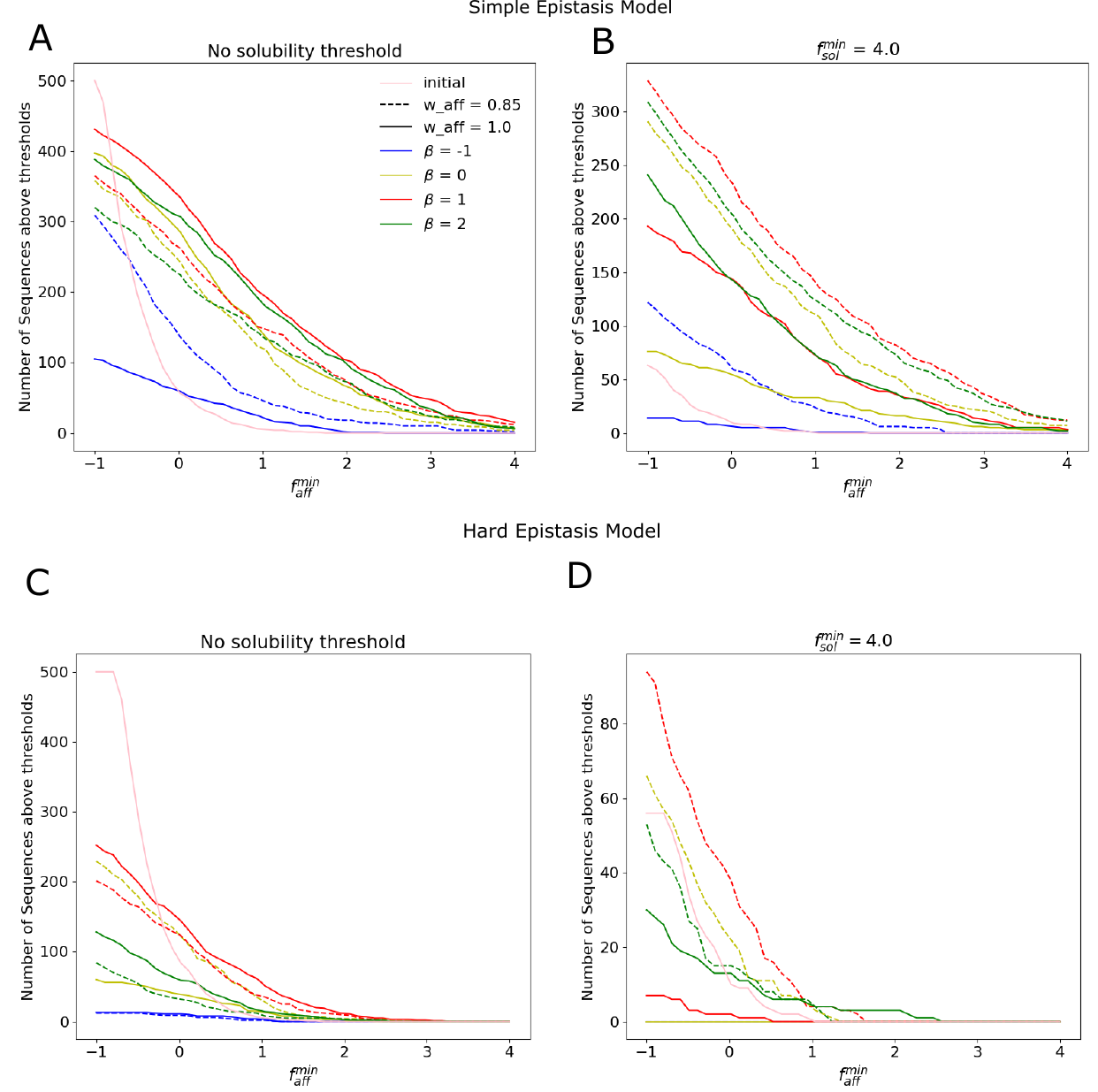}
\end{center}
\caption{
Figure A. shows the number of sequences out the top 500 selected from the set of generated sequences with Metropolis Hastings when trying to optimize the simple epistasis model with an affinity above a certain thresholds for different choices of $\beta$ and $w$. We compare those results to the set of sequences in the training set, indicated as initial. B shows the percentage of sequences that are above a certain threshold and also have a predicted solubility score above 3. Figure C. and D. show the results when trying to optimize the hard epistasis model.
}
\label{fig:4}
\end{figure*}

{\PP Using the synthetic dataset we investigate the most appropriate choice of $\beta$ for lead optimization. We  observe that for three of the tasks: simple with and without a solubility threshold and hard without a solubility threshold, the best choice of $\beta$ out of the ones we tested is 1.0. In the case where we have no solubility threshold (Figure \ref{fig:4} A,C), the best choice is for $w = 1.0$ whereas in the simple case with $f_{\rm sol}^{\rm min} = 4.0$, $w = 0.85$ (Figure \ref{fig:4} B). These results suggests that using an optimistic choice of $\beta$ helps to discover more optimal sequences rather than a pessimistic choice of $\beta = -1.0$. In fact, in the considered scenarios, using $\beta = -1.0$ consistently performs worse than the other choices.

In addition, we observe that being too optimistic can be detrimental as $\beta = 2.0$ can perform worse than $\beta = 1.0$. For the hard task with no solubility threshold using $\beta= 1.0$ yields approximately twice as many valid sequences for all affinity thresholds. However, for the simple task $\beta = 2.0$ and $\beta = 1.0$ perform approximately the same, since in the hard task finding a beneficial mutant randomly is less likely than in the simple task. 
Furthermore, we observe that for the most difficult case, the hard task with a solubility threshold, using $\beta = 1.0$ is also the best choice, except for $f_{\rm aff}^{\rm min} > 1$, when  $\beta = 2.0$ and $w = 1.0$ is best. We found that using an optimistic choice for $\beta$ yields consistently better results than a pessimistic choice of $\beta$, even on the hard task. In fact, the more difficult the task, the better it is to use $\beta = 2.0$ rather than $\beta = 1.0$. This suggests that in order to find the few sequences that are optimized for solubility and affinity, it is necessary to take more risks during the exploration process.}

{\HM{To explore the consequences of small budgets, we considered $B = 20$ (Fig \ref{figsi:9}) and found for the hard task with  $f_{\rm sol}^{\rm min} = 4.0$, the best choice is to use $\beta = 0$ and $w = 0.85$. This suggests that when having a limited budget and optimizing for a difficult task, it may be better to use a more conservative choice for $\beta$ and generate high confidence sequences. }}

\subsection*{Comparison to other methods}

We can use the synthetic dataset to compare the performance of our approach to previous methods against the ground truth.
Khan et al~~\cite{khan2022antbo} define the multi-objective optimization of antibodies as a constrained affinity optimization problem. They use a local search algorithm similar to hill climbing as the inner loop of an active learning framework to optimize the CDRH3 of antibody heavy chain in order to increase their affinity according to a synthetic affinity function computed with the software Absolut!~\cite{robert2022unconstrained}, as well as three developability properties (net charge, repetition of amino acids and presence of a glycosylation motif). 
{\PP For each developability property, the algorithm requires an interval of valid values to be specified. Sequences that fall outside of these developable regions are automatically rejected by the search algorithm.} 

In order to compare the local search algorithm from antBO to our approach, we implemented their method and ran it using two different sets of developability restrictions. In the first case we limit the exploration to sequences with Hamming distance less than $6$ to the wild type. In the second case, we add the additional restriction that the sequences must have a solubility predicted score $\hat f_{\rm sol} \geq 4$. For both cases, we ran the local search algorithm $800$ times in parallel for $200$ steps and kept the last sequences of each run as the set of generated sequences.

In addition, as a negative control, we also compared our method to a set of $500$ randomly generated sequences $D_{\rm rnd}$. We generated these sequences by starting with the best mutant in $D_{\rm syn}$. We then randomly select $3$ positions that have no yet been mutated. For each of these positions, we perform an amino acid substitution with an amino acid uniformly drawn from the $19$ other amino acids.

\begin{figure*}
\begin{center}
  \includegraphics[width=0.8\linewidth]{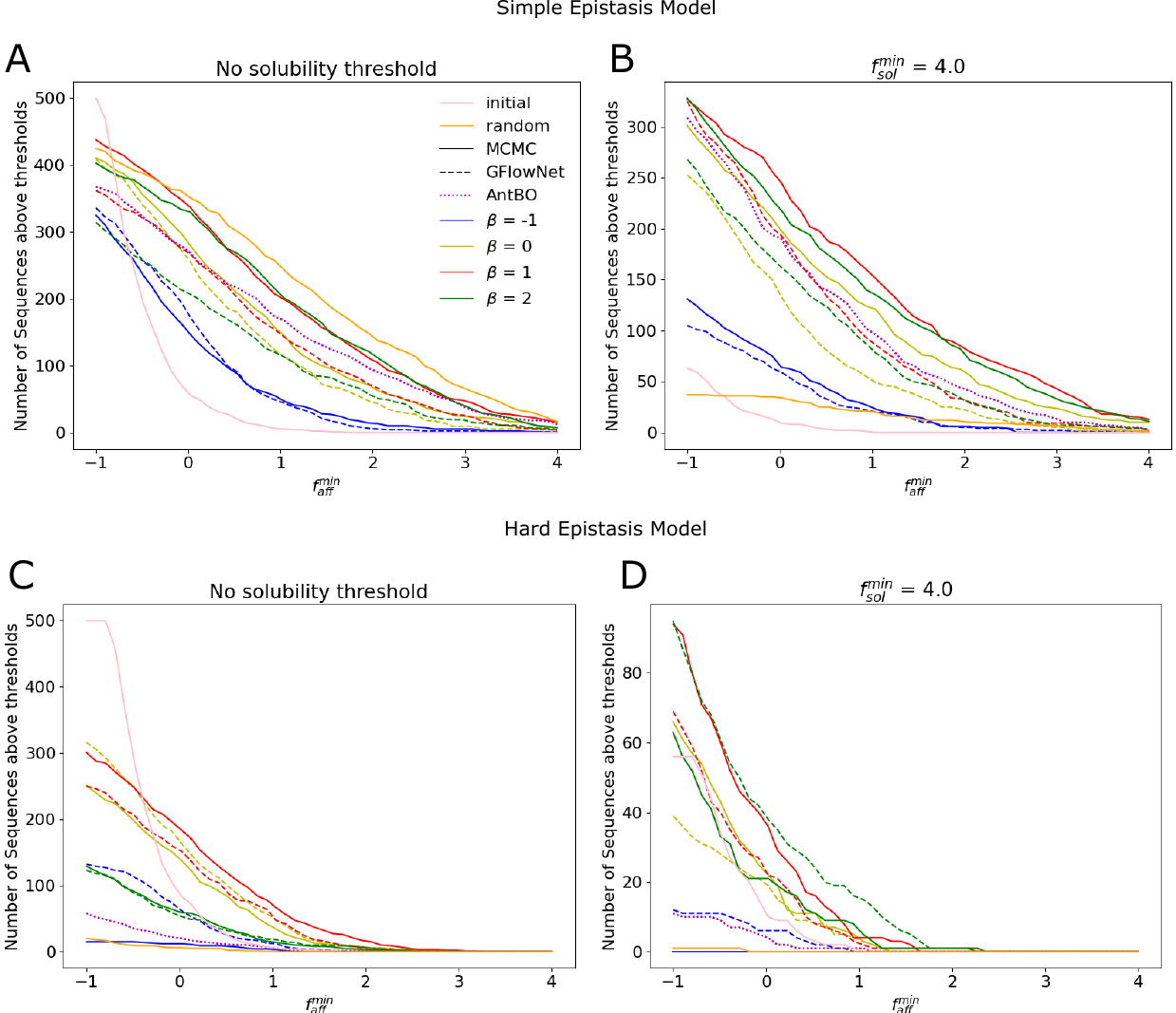}
\end{center}
\caption{
Figure A shows the number of sequences out the top 500 selected from the set of sequences generated for different choices of $\beta$ and generative method (Metropolis Hastings and GFlowNet). We compare those results to the set of sequences generated by the local search procedure of antBO and the set of sequences used to train the Gaussian Process affinity predictor, indicated as initial, as well as a set of randomly generated sequences in the neighborhood of the best sequence in $D_{\rm rnd}$ indicated as random. B shows the percentage of sequences that are above a certain threshold and also have a predicted solubility score above 3. Figure C and D show the results when trying to optimize the hard epistasis model.
}
\label{fig:5}
\end{figure*}

We compared the initial mutants $D_{\rm syn}$, the randomly generated mutants $D_{\rm rnd}$, and the sequences generated with the local search of antBO to sequences generated with our method using MCMC sampling and the GflowNet (Fig.~\ref{fig:5}). For both MCMC and the GflowNet, for each choice of $\beta$ we combined all the sequences generated for both choices of the affinity weight $w$ into a single set. For each set of generated sequences, we remove all sequences with $f_{\theta}^{\rm sol}(x) < f_{\rm sol}^{\rm min}$ for $f_{\rm sol}^{\rm min} = -\infty$ and $f_{\rm sol}^{\rm min} = 4.0$. Then we select the top $500$ sequences according to the acquisition function used during the generation process. We then compute the percentage of sequences with a synthetic affinity $f{\rm aff}$ higher than a chosen threshold $f_{\rm aff}^{\rm min}$. 

 % fig5

For the simple task with no solubility threshold, MCMC sampling with $\beta = 1$ and $\beta = 2$ are the best choices of parameters and outperform the antBO local search (Fig.~\ref{fig:5} A). However, we observe that the random sequences contain more optimized sequences than any of the sequences generated by our method. This suggests that the task is quite easy. For the simple task with $f_{\rm sol}^{\rm min} = 4.0$, MCMC sampling with $\beta = 0$, $\beta = 1.0$ and $\beta = 2.0$ outperforms the antBO local search (Fig.~\ref{fig:5} B). In addition, MCMC sampling performs better than GFlowNet for similar values of $\beta$ except for $\beta = -1.0$ where the results are similar. All of the methods do better than the random baseline and the initial dataset. For the simple task, regardless of solubility thresholds, we observe that using a positive $\beta$ is more beneficial although if $\beta$ is too high, the performance starts to decrease a little.

For the hard task with no solubility threshold, the best method is MCMC sampling  with $\beta = 1.0$ (Fig.~\ref{fig:5} C). In this case, using $\beta = 2.0$ leads to significantly worse performance. In addition MCMC sampling with $\beta = 1.0$ outperforms the GflowNet, except for $\beta = 0.0$ for which they give similar results. Finally both MCMC sampling and Gflownet outperform the local search of antBO, the initial dataset and the random baseline for $f_{\rm aff}^{\rm min} > -0.3$. 
Finally, on the hard task with $f_{\rm sol}^{\rm min} = 4.0$, we find that for an affinity threshold $f_{\rm aff}^{\rm min} > 0$, the GFlowNet with $\beta = 2.0$ generates the highest number of valid sequences. This is the first setting where the GFlowNet performs better than MCMC sampling. Overall, both MCMC sampling and the GFlowNet outperform the local search of antBO, the initial dataset and the random baseline.

{\HM Comparing the methods for a limited budget of $B = 20$ (Fig.~\ref{figsi10}), we find the most significant differences for the hard task with $f_{\rm sol}^{\rm min} = 4.0$, where using $\beta$ = 0 greatly outperforms using $\beta = 2$. This result suggests conservative choices of $\beta$ for limited budgets. Furthermore, on all tasks except the simple one with no solubility threshold, our energy based method outperforms the local search of antBO. On the simple task with no solubility threshold, even antBO performs worse than random.}

These results  serve to demonstrate that our energy based sampling method performs better than constrained optimization in a variety of settings. In particular, we find that the local search algorithm has more difficulty when generating sequences optimized for the hard epistasis affinity function than for the simple epistasis affinity function.

\section{Discussion}

 % discussion
\subsection*{Data limitations and choice of models}
{\PP We made many design choices  when developing this method. First and foremost, data limitations  are inherent to the task of antibody optimization. In this paper, we use a dataset that contains $10^5$ sequences. While this number may vary depending on the available resources, that number is usually a tiny fraction of the space of sequences we wish to explore. Even  limiting the search to sequences 6 amino acids away from the WT, leaves $7.088 \times 10^{13}$ possible candidates. Given the complexity of the task of predicting the affinity of an antibody to a target antigen based on sequence information alone,  training a model on the dataset that  generalizes well to the entire search space is hard.} 
For this reason, we chose to use a Gaussian Process as this method allows us to directly estimate the epistemic uncertainty of our prediction. Furthermore, due to the limited training set size, when using a protein language model to compute the embeddings of sequences, we chose not to do any fine-tunning of the model to avoid potential over-fitting due to the large number of parameters these models use.

For the solubility model, in absence of a large exploitable dataset for machine learning, we opted to reuse the method developed by \cite{jain2017prediction}. Since their method relies on first training a SASA prediction method on publicly available structures of antibodies, we were able to implement our own SASA prediction model and then reuse the 20 hydrophobicity weights learned by \cite{jain2017prediction} on their private dataset containing 5000 pairs of antibody sequences and their HIC RT.

\subsection*{Review of prior approaches}

{\PP A number of methods for the generation of optimized antibody sequences have previously been proposed. Like our method, AntB0 \cite{khan2022antbo} performs multi-objective optimization by modifying the heavy chain of antibodies. They seek to optimize the affinity estimate given by Absolut! as well as three developability properties. Another similar  approach \cite{zeng2024antibody}. also performs multi-objective optimization and optimizes both affinity and thermostability.
Our approach to multi-objective optimization is different from Ref. \cite{zeng2024antibody} and Ref. \cite{khan2022antbo}. Both of these methods define hard constraints for the developability properties. For example, antBO requires the charge of the generated sequences to be between -2 and 2. The method in Ref. \cite{zeng2024antibody} requires the predicted melting temperature to be above 60 or 65 $^\circ$C.
We do not use hard constraints but instead use an energy function made up of a linear combination of the predicted properties and sampled from the Boltzmann distribution defined by this energy function. The advantage of this approach is that it does not require prior knowledge of constraints for each property. In the case of the solubility prediction for example, the HIC RT can depend on experimental conditions and defining a precise threshold for which antibodies are soluble or not is often not simple. 
Ref. \cite{gessner2024active} developed an active learning loop procedure to optimize the affinity of antibodies based on the predicted free energy of the antibody antigen complex using the Schr\"odinger software, using a standard multi-round active learning framework without constraining the search space. Unlike us, they do not perform multi-objective optimization but they use a gaussian process as a predictive model for the Schr\"odinger $\Delta\Delta G$ prediction similar to our methods and \cite{zeng2024antibody} and Ref. \cite{khan2022antbo}.
These three  methods \cite{gessner2024active, zeng2024antibody, khan2022antbo} use the expected improvement acquisition function to select which sequence to test. In contrast, we use the UCB/ LCB acquisition function in order to study the effect of being optimistic vs pessimistic in a single round of optimization. Another key difference is that, we utilize a pretrained autoregressive protein language model to regularize our generative process. Of these methods, Ref \cite{zeng2024antibody} is the most similar to us in scope as it is designed to perform offline single round multi-objective optimization, although it optimizes thermostability whereas we optimize solubility. In addition, the idea of performing the exploration in the latent space of the auto-encoder in order to amortize the exploration cost of the search space is similar to our use of the GflowNet. antBO and Ref. \cite{gessner2024active}  on the other hand were designed for multi-round online active learning. Both methods use in-silico affinity prediction methods as $f_{\rm aff}$, allowing them to bypass the more expensive and time-consuming wet lab experiments.}

In addition, \cite{li2023machine} trained a GP on the dataset of \cite{engelhart2022dataset} and generated new sequences using Monte Carlo Markov chain and the expected improvement acquisition function. They demonstrated that this approach is able to generate new mutants with higher affinity to the CB-119 peptide than in the initial dataset. Unlike us however they do not perform multi-objective optimization.

\subsection*{Motivation behind investigating the choice of acquisition function}

Our motivation to investigate the use of different choices of $\beta$ for the acquisition function came from the fact that we have a limited dataset and the need to constrain the space of sequences on which the exploration is performed, both to facilitate the generative process and limit the search to sequences on which our predictive models are confident. By varying $\beta$, we can choose to let the generative model generate sequences farther away from the original dataset or to remain close. In \cite{zeng2024antibody} \cite{khan2022antbo}, a trust region is used to achieve a similar goal. The trust region limits the exploration and is dynamically updated at the end of each optimization round based on whether a new best solution has been found. It is not possible to use the same approach in the context of a single round as the trust region is only modified at the end of each round. In our case however, trying out different choices for $\beta$ is a principled way of restricting the search space.

We took inspiration from recent papers in offline reinforcement learning to try both optimistic and pessimistic values of $\beta$ \cite{moskovitz2021tactical}\cite{NEURIPS2021_34f98c7c}. The setting of offline reinforcement learning is identical to single round antibody optimization, wherein we do not have the ability to further query the environment or run more wet lab experiments. In this setting, multiple papers have pointed out that it may be useful to penalize uncertainty, i.e. choose a negative value for $\beta$ \cite{shi2022pessimistic}\cite{koppel2024information}.

The advantage of being pessimistic is that it forces the generative model to generate sequences on which it is more confident that its prediction is accurate. This is particularly important if the number of sequences that can be tested is small and if there are few good sequences (with a high affinity) within the trust region. Instead of being optimistic and generating a large variety of sequences for which we have no guarantee that they will work, we chose to generate sequences closer to the ones in our dataset, but for which we have more accurate affinity predictions, and therefore stronger guarantees that they will bind to the target antigen.

{\HM However, we found that risk was often necessary to find new mutations. The only exception was when using a limited budget on the most difficult task(Fig \ref{figsi:9}). Yet even in that case,  $\beta = 0$ yielded better candidates than $\beta = -1$.}

\subsection*{Limitations of our method and future research directions}

There are several limitations to our study. We only consider the case of optimizing over both affinity and solubility, although there are several other important properties of interest, such as thermostability. It will be interesting to test this method when there are multiple properties to optimize over.

Our solubility predictive model only outputs a predictive score but no uncertainty estimate. It remains  a possible research direction to build a SASA prediction model that includes uncertainty estimates. In addition, since the model outputs a prediction for every amino acid residue, it would also need to output an uncertainty estimate for every residue. It would therefore also be interesting to investigate how to combine each estimate into a single uncertainty score that could be included in the acquisition function.

{\PP Overall, our approach demonstrates the possibility to optimize several antibody properties in parallel. As real-life pharmaceutical projects aims at optimizing more than only 2 protein properties (affinity ans solubility) before identifying a drug candidate, a direct application of this work would be to apply this approach to the optimization of 5 or more properties with experimental evaluations of the resulting sequences. By providing sequences fulfilling the drug developability criteria, the method described in this paper can help scientists to accelerate drug discovery projects }

\begin{comment}
Finally, generative models can be integrated inside an active learning loop as a way to recommend new sequences to be tested. In an active learning framework, sequences generated at some active learning round $t$ are evaluated in-vitro or during clinical-trials. The results of these experiments are used to update the generative model after which a new round of hopefully better drug candidates are generated. This process can be repeated until a functional drug has been found.

The Engelhart and al.~\cite{engelhart2022dataset} dataset is representative is ideal for a lead optimization task as it contains  mutants of the WT sequence  we wish to improve. In addition, we know that all antibodies bind to the same epitope. This is not always the case when the antigen of interest is a large protein which can complicate the training of the predictive model. The number of evaluated mutants is also of the same order of magnitude than can be produced during a lead optimization project.
\end{comment}

\section{Methods}

 % methods
\subsection{Solubility predictor}\label{sec:solpred}

In order to select the structures from which to build the training dataset for the solubility predictor, we took all the structures available in SabDab~\cite{dunbar2014sabdab} and clustered them by 98\% similarity of the heavy chain sequence using CD-HIT~\cite{fu2012cd}. From each cluster, we selected the antibody with the longest heavy chain sequence, giving us 2648 distinct heavy chain structures. We use the Shrake-Rupley algorithm~\cite{shrake1973environment} to compute the SASA of every residue in the heavy chains. We divided each computed SASA value by the maximum exposed side-chain SASA in the Ala-X-Ala peptides (where X is the amino acid for which the SASA was computed) as determined by~\cite{chennamsetty2010prediction}.

We use the convolutional neural network architecture used for NanoNet~\cite{cohen2022nanonet} with an additional embedding layer added at the beginning. This layer takes as input a one-hot vector encoding of dimension $L \times 22$. The first 20 values are used to determine which amino acid is present at position $i$ and the last two values contain the distance to the first amino acid $i$ and the distance to the last amino acid $L-i$ ~\cite{sethna2019olga}. We found that providing both the forward and backward positioning information helps the network deal with sequences of different lengths. The output of the encoding layer is a matrix of dimension $L \times h$ with $h = 64$.

The network was trained using stochastic gradient descent for 10 epochs and using a mini-batch size of 16. We use the MSE loss function:
\begin{equation}
    L_{\theta} = \frac{1}{L} \sum_{i = 1}^{L} ({\rm SASA}(j,x_j;\theta) - {\rm SASA}_{\rm struct}(j,x_j;\theta))^2,
\end{equation}
where ${\rm SASA}(j,x_j;\theta)$ is the predicted SASA of the residue at position $i$ in the sequence $x$ and  $ {\rm SASA}_{\rm struct}(j,x_j;\theta)$ is its true value.

\subsection{Gaussian Process} \label{sec:GPexp}

The Gaussian process assumes that the output function $f(x)$ was initially drawn at random, but in a way that similar $x$ have similar $f(x)$. The distribution $f(x)$ for all $x$ is assumed to be a multivariate Gaussian of constant mean $C$ and covariance $k(x,x')$, where $k$ is called the kernel function and determines how similar the outputs of $x$ and $x'$ should be.

Let $X \in \mathbf{R}^{n \times m}$ be the set of training data and $Y \in \mathbf{R}^{n}$ associated affinity values. We assume that the data is noisy so that $y(x) = f(x) + \epsilon(x)$ with $\epsilon(x) \sim N(0,\sigma_n^2)$.
Call $z$ is a new sequence for which we aim to make a prediction for $f(z)$.

The joint distribution for the training data and the new sequence is also a multivariate Gaussian: $p(Y,f(z)|X,z) \sim N(\mu, \Sigma)$ with:
\begin{equation}
\mu = \begin{pmatrix}
C\\
C\\
...\\
C
\end{pmatrix}
 , \; \;
 \Sigma = \begin{pmatrix}
k(X,X) + \sigma_n^2I_n & k(z,X)\\
k(z,X)^T & k(z,z)
\end{pmatrix}.
\end{equation}
$k(X,X)$ a matrix of dimensions $n \times n$ where the $(i,j)$ entry is $k(x_i,x_j)$. $k(z,X)$ a vector of dimension $n$ where the ith entry is equal to $k(z,x_i)$ and $I_n$ is the identity matrix of dimension $n$.

We seek the posterior distribution: \begin{equation}p(f(z)|X,Y,z) = \frac{p(Y,f(z)|X,z)}{p(Y|X)}\end{equation}
which is Gaussian with mean:
\begin{equation} \mu(f(z)|X,Y) = C+ k(z,X)^T(K + \sigma_n^2I_n)^{-1}(Y - C),\end{equation}
and variance
\begin{equation} \sigma^2(f(z)|X,Y,z) = k(z,z) + k(z,X)^T(K + \sigma_n^2I_n)^{-1}k(z,X).\end{equation}
The posterior distribution for the noisy prediction $y(z)=f(z)+\epsilon(z)$ has an additional $\sigma_n^2$ term in its variance.

The model provides both an estimate for the affinity of a new sequence as well as an uncertainty estimate. This uncertainty $\sigma$ only depends on the training data $X$ and can be reduced by adding more training examples. Therefore it captures the ``epistemic'' uncertainty of the error. On the other hand $\sigma_n$ is independent of the training set and $z$, therefore it captures the ``aleatoric'' uncertainty.

We use the common radial basis function (RBF) Kernel:
\begin{equation}\label{eq:RBF}
    k(x,x') = \delta \exp(-\frac{\Vert x - x'\Vert^2}{2\lambda^2}).
\end{equation}
The RBF kernel~\cite{williams2006gaussian} encodes the bias that sequences close to one another in embedding space have similar affinities to the target antigen.

The norm $\Vert x-x'\Vert$ is defined as the Euclidian distance in the embedding space. We tested the following embedding choices:
\begin{itemize}
    \item One-hot vector embedding of the amino acid sequences. For a sequence of length L, this a vector of length $L \times 20$.
    \item Embeddings computed by the protein language model ESM2~\cite{lin2022language}. ESM2 is a BERT~\cite{devlin2018bert} style transformer that was trained using the masked language task on general proteins. We compared the versions with 8M, 150M and 650M parameters.
    \item Embeddings computed by Antiberty~\cite{ruffolo2021deciphering}, a bert style transformer, similar to ESM2, but trained uniquely on human antibody heavy chain sequences.
\end{itemize}

The parameters $\lambda$ and $\delta$ of the Kernel $k(x,x')$, as well as $\sigma_n$, are learned by minimizing the log marginal likelihood of the training data:
\begin{multline}\label{eq:lml}
\ln p(Y|X) = -\frac{1}{2}Y^T(K + \sigma_n^2I)^{-1}Y - \frac{1}{2} \ln 2\pi|K + \sigma_n^2I|.
\end{multline}

\subsection{Humaness Model}

To model humaness parameter $p_{\rm HUM}$, we use  IGLM~\cite{shuai2021generative}, an auto-regressive transformer trained on 558M human heavy chain sequences from the OAS database~\cite{olsen2022observed}. 
Given the amino acid sequence from position 1 to $i-1$ denoted $x_{<i}$, the model returns a discrete probability distribution over the set of the 20 amino acids $p_{\text{IGLM}}(x_i | x_{<i})$ and adds the next amino acid. 
For each sequence of amino acids $p_{HUM}$ is the probability that IGLM generates that particular sequence (obtained using the likelihood method of IGLM)
\begin{equation}
    p_{\rm HUM}(x) = \prod_{i=1}^n  p_{\text{IGLM}}(x_i | x_{<i}),
\end{equation}
which guarantees that only amino acid sequences that resemble human heavy chain sequences will have a high probability of being generated.
Shuai and al~\cite{shuai2021generative} showed that sequences generated with IGLM have on average good developability properties when looking at solubility, aggregation and CDRH3 length.

\subsection{Energy based model}

We sample sequences using Energy based models (EBMs) by assigning an energy score $E(x)$ to each amino acid sequence $x$ where $E(x)$ is low for desirable $x$.
We would like to sample $x$ from a probability distribution that is as close as possible to the distribution of natural antibodies, $p_{\rm HUM}$, while minimizing the mean of $E(x)$:
\begin{equation}
\label{eqn:2}
    p = \argmax_{\pi \in \Pi} (\sum_x \pi(x)E(x) + T D_{\rm KL}(\pi||p_{\rm HUM}),
  \end{equation}
  where $D_{\rm KL}$ is the Kullback-Leibler divergence
  The solution of this minimization is given by the Boltzmann law, Eq.~\ref{eq:bolz}. In the particular case where $p_{\rm HUM}$ is replaced by the uniform distribution, $D_{\rm KL}$ becomes the negative entropy of $\pi$, and the problem reduces to the maximum entropy principle, where $T$ plays the role of an inverse Lagrange parameter enforcing the mean value of $E(x)$.

The temperature $T$ sets the trade-off between the objectives of minimizing the quantity of interest $E(x)$, and remaining as close to the basal distribution of antibodies. While $T=0$ reduces to finding the best sequences, $T>0$ ensures a higher diversity of antibodies that look like natural ones.

The energy function $E(x)$ is chosen as a linear combination of multiple properties which we wish to minimize, called linear scalarization:
\begin{equation}
    E(x) = \sum_i w_i \, f_i(x) \; \; \text{s.t} \; \; \sum_i w_i = 1.
\end{equation}
Varying the weights $w_i$ make it possible to explore the Pareto front in the $T=0$ limit, when that front is convex. For $T>0$, the vicinity of the front is explored while also ensuring that antibodies still look like natural ones drawn from $p_{\rm HUM}$.

\subsection{MCMC sampling}

To perform MCMC sampling of Eq.~\ref{eq:bolz}, we start from the WT sequence 'GFTLNSYGISIYSDGRRTFYGDSVGRAAGTFDS', the concatenation of the CDRH1, CDRH2 and CDRH3 of the AB-14 sequence from the CB-119 dataset. The same wild type sequence was also use to generated the synthetic datasets for the simple and hard synthetic affinity task. We set this sequence to be the first  sequence $x_0$. 
At time $k$, we randomly sample a mutant sequence $x'$ from the neighborhood of $x_k$. The neighborhood of $x_k$ is defined as the set of sequences that are at most 1 mutation away from $x_k$ and at most $6$ mutations away from $x_0$. For each sequence with probability ${\rm min}(1,p(x')/p(x_k))$, we accept the mutation and set $x_{k+1} = x'$, or otherwise we reject the mutation and set $x_{k+1} = x_k$.

We ran this process in parallel 8 times. Each time we performed MCMC sampling for a period of 20000 time steps.
To add a  burn-in period,  we removed all the sequences $x_1, x_2, ... x_b$ from the 8 chains and computed the Gelman-Rubin statistic. {\PP The Gelman-Rubin statistic checks the convergence of multiple Markov Chain Monte Carlo (MCMC) chains by computing the ratio between the variance within each chain to the variance between chains, with values close to 1 indicating convergence}. We then selected an initial timestep $b$ such that the Gelman-Rubin statistic was as close to 1 as possible. If we could not find a timestep $b$ such that the Gelman-Rubin statistic was lower than 1.1, we ran the sampling for 20000 additional steps and repeated the process until convergence.

\subsection{GFlowNet sampling}

For the autoregressive model used in our implementation of the GFlowNet, we used a ByteNet \cite{bytenet} \cite{bytenetcomp} architecture. The network starts with an encoding layer that feeds into 4 ByteNet blocks. Each block uses masked convolutional layers with no dilution and a kernel size of 17. The output of these 4 blocks is a matrix of dimension $L \times H$ where $L$ is the maximum sequence length and $H$ is a hyper parameter representing the size of the encoding which we choose to be 16. We then feed to a decoder the $i^{th}$ column of this matrix to which we append two values, $i$ and $n$, where $i$ is the position of the amino acid we are currently sampling and $n$ is the Levenshtein distance between the sequence generated so far and the prefix of length $i-1$ of the WT sequence. We provide $i$ to the decoder to simplify the learning process and $n$ in order for the network to be able to learn to only generate sequences with at most $6$ mutations from the wild type sequence.

The decoder consists of a 2 layer multi-layer perceptron with a normalizing layer and ReLu activation layer between the first and second linear layer. The output of the second layer is vector of dimension 20 containing the logits for each of the 20 amino acids.

For each GflowNet trained, we use a starting learning rate for the parameters of the network of $10^{-3}$, except for the learning rate of $Z_\theta$ which is set to $10^{-2}$ (an order of magnitude higher per the recommendation of the GflowNet paper). We train each network for 8000 training steps, halve the learning rate after 4000 steps and use a replay buffer with a max size of 20000. We initialize to contain the sequences in the COVID dataset for the COVID task and the synthetic dataset for the simple and hard synthetic tasks. At each step, we first generate 16 sequences using the generative model and compute their energy $E$. We then sample 16 more sequences and their score from the replay buffer and update the parameters of the model using stochastic gradient descent on the set of 32 sequences. The 16 generated sequences are then added to the replay buffer and the process is repeated.

Unlike for the MCMC, there is no principled way to check that the training of the neural network has converged to a local optimum. In order to verify that the network is learning to properly generate sequences with a probability proportional to their reward, we take inspiration from Ref.~\cite{madan2023learning}
Using a separate dataset of sequences from the sequences used to initiate the replay buffer, we periodically, during the training process compute the log probability of the sequences being generated. We compute the SpearmanR correlation score between the scores and the log probability of generated sequences.

While this method has its flaws, we found it was a simple method to verify that the generative model was correctly learning for each choice of $\beta$, $w$ and inverse temperature $T^{-1}$ (SI Fig.~\ref{figsi2}).

\section*{Data availibility}
The code and data supporting the findings of this study are available at the following GitHub repository:
https://github.com/statbiophys/ABGen

\section*{Acknowledgements}
This work was supported by Sanofi, the European Research Council
consolidator grant no 724208 (AMW, TM), and the Agence Nationale
de la Recherche grant no ANR-19-CE45-0018 “RESP-REP” (AMW, TM,
PP). PP and HM are Sanofi employees and may hold shares and/or stock options in the company.
The authors declare that this study received funding from Sanofi. The funder collaborated directly in the study and was involved in the study design, analysis, and interpretation of data, the writing of this article, and the decision to submit it for publication.

\bibliographystyle{pnas}

\onecolumngrid

\newpage

\section*{Supplementary information}

 % si

This section provides supplementary figures that support the main findings described in the document.

\renewcommand{\thefigure}{S\arabic{figure}}
\setcounter{figure}{0} 
 % figsi1
\begin{figure}[H]
\begin{center}
  \includegraphics[width=\linewidth]{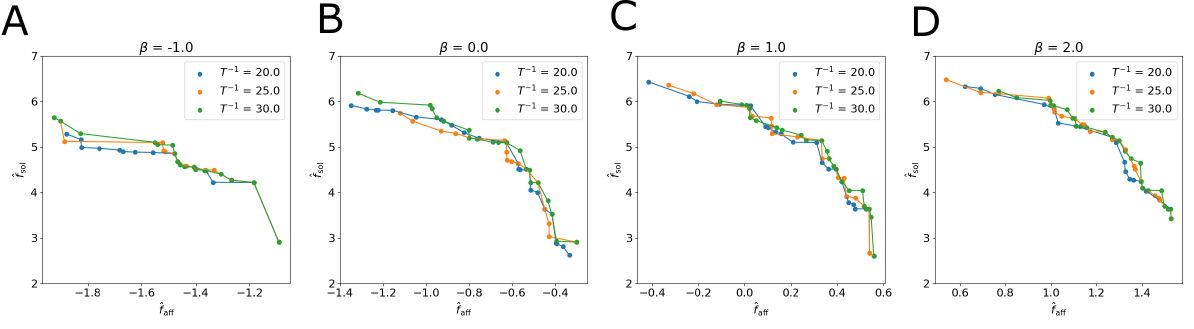}
\end{center}
\caption{A. Empirical Pareto Front on the Sars-Cov-2 task for $\beta = -1$ and $T^{-1} \in [20.0,25.0,30.0]$ B. Empirical Pareto Front on the Sars-Cov-2 task for $\beta = 0$ and $T^{-1} \in [20.0,25.0,30.0]$ C. Empirical Pareto Front on the Sars-Cov-2 task for $\beta = 1$ and $T^{-1} \in [20.0,25.0,30.0]$ D. Empirical Pareto Front on the Sars-Cov-2 task for $\beta = 2$ and $T^{-1} \in [20.0,25.0,30.0]$}
\label{figsi1}
\end{figure}

 % figsi2
\begin{figure}[H]
\begin{center}
  \includegraphics[width=\linewidth]{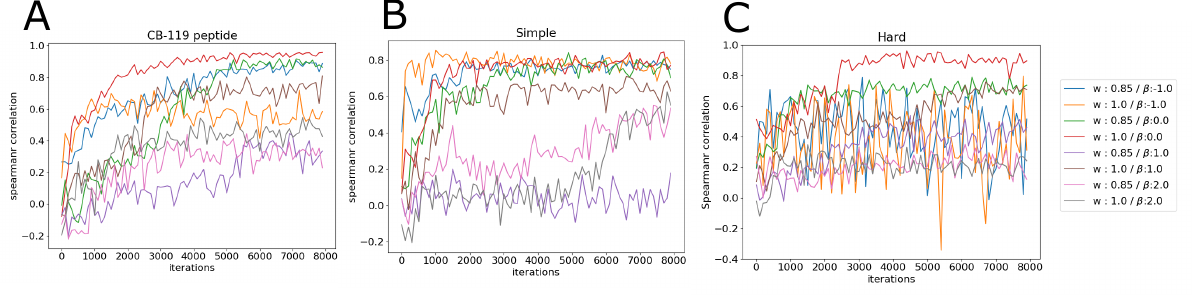}
\end{center}
\caption{During the training of the GflowNet, for each task (Sars-Cov-2/Simple/Hard) and hyper-parameter choice ($\beta$ / $w$ / $T^{-1}$, we randomly selected 128 sequences generated by Metropolis Hastings on the same task and for the same hyper-parameter choices and computed the spearman correlation score between the log probability of the sequences being generated and their score $R(x) = {\rm} ln  \, p_{\text{HUM}} - E(x)$. A. Shows the results on the Sars-Cov-2 task for $T^{-1} = 20.0$, $w \in [0.85,1.0]$ and $\beta \in [-1.0,0.0,1.0,2.0]$, B. Shows the results on the simple task for $T^{-1} = 10.0$, $w \in [0.85,1.0]$ and $\beta \in [-1.0,0.0,1.0,2.0]$ and C. shows the results on the hard task for for $T^{-1} = 10.0$, $w \in [0.85,1.0]$ and $\beta \in [-1.0,0.0,1.0,2.0]$. We can observe that the Gflownet training converges faster for lower values of $\beta$}
\label{figsi2}
\end{figure}

 % figsi3
\begin{figure}[H]
\begin{center}
  \includegraphics[width=\linewidth]{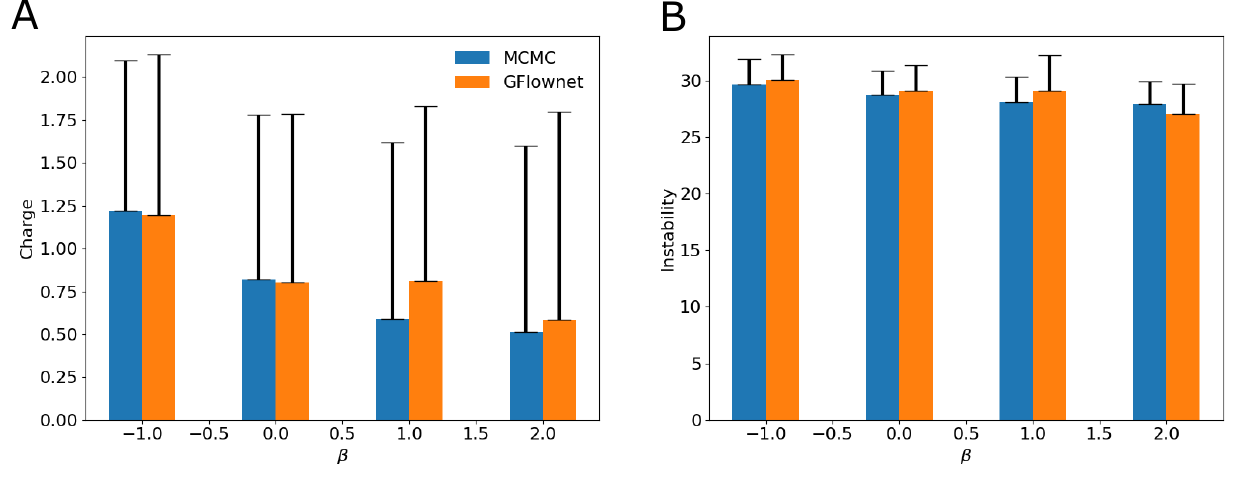}
\end{center}
\caption{Figure A. shows the average and the standard deviation of the charge of the heavy chain sequences generated by Metropolis Hastings and the GFlowNet for different choices of $\beta$. A net charge between -2 and 2 is considered desirable.Figure B. shows the average and the standard deviation of the instability scores of the heavy chain sequences generated by Metropolis Hastings and the GFlowNet for different choices of $\beta$. An instability score lower than 40 is desirable.}
\label{figsi3}
\end{figure}

 % figsi4
\begin{figure}[H]
\begin{center}
  \includegraphics[width=\linewidth]{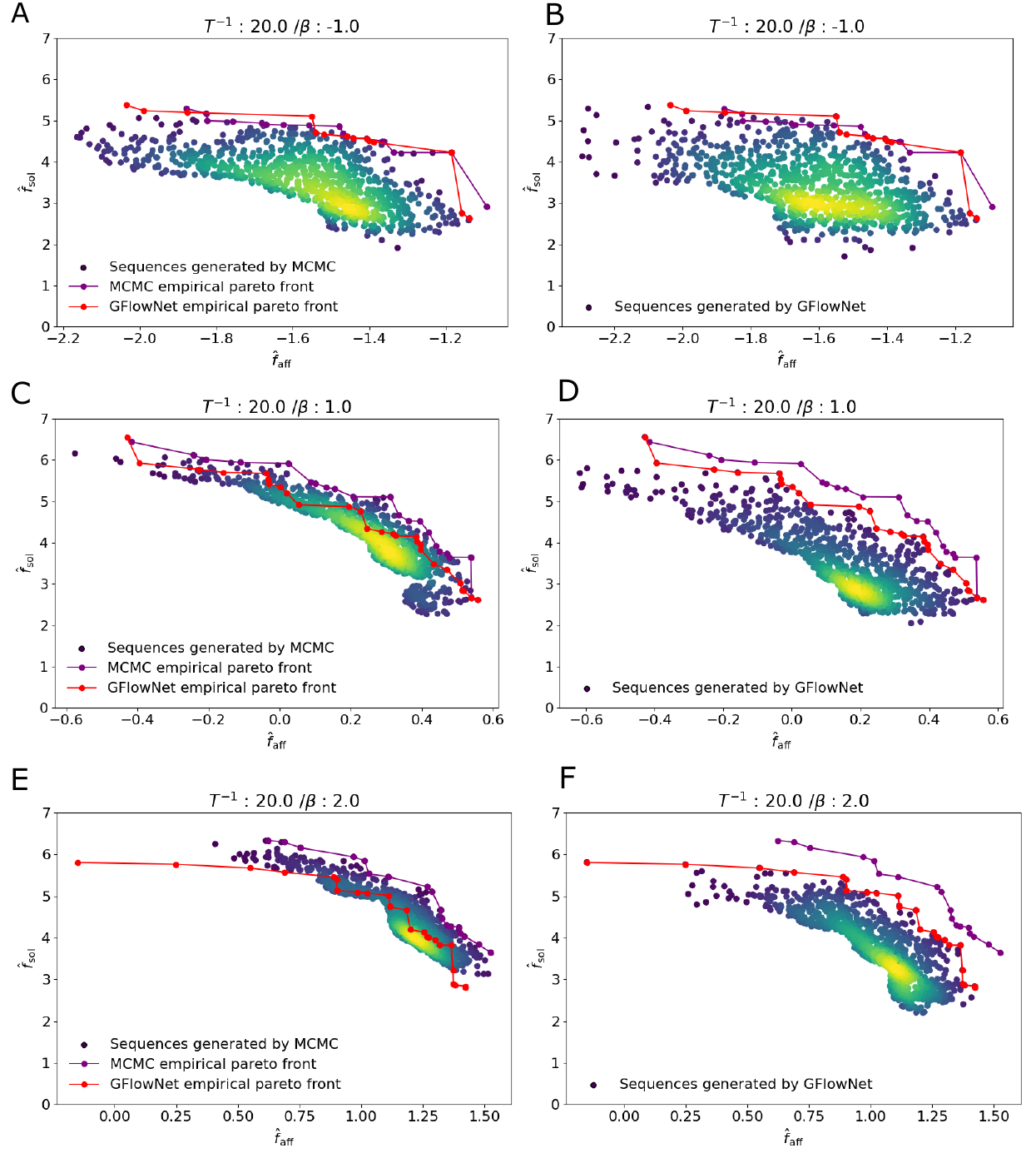}
\end{center}
\caption{A. Density plots of 1000 sequences sub-sampled from the set generated by Metropolis-Hastings at inverse temperature $T^{-1} = 20$ and $\beta = -1.0$. B. Density plots of 1000 sequences sub-sampled from the set generated by the GFlowNet at inverse temperature $T^{-1} = 20$ and $\beta = -1.0$ C. Density plots of 1000 sequences sub-sampled from the set generated by Metropolis-Hastings at inverse temperature $T^{-1} = 20$ and $\beta = 1.0$. D. Density plots of 1000 sequences sub-sampled from the set generated by the GFlowNet at inverse temperature $T^{-1} = 20$ and $\beta = 1.0$,  E. Density plots of 1000 sequences sub-sampled from the set generated by the Metropolis Hastings at inverse temperature $T^{-1} = 20$ and $\beta = 2.0$,  F. Density plots of 1000 sequences sub-sampled from the set generated by the GFlowNet at inverse temperature $T^{-1} = 20$ and $\beta = 2.0$}
\label{figsi4}
\end{figure}

 % figsi6
\begin{figure}[H]
\begin{center}
  \includegraphics[width=\linewidth]{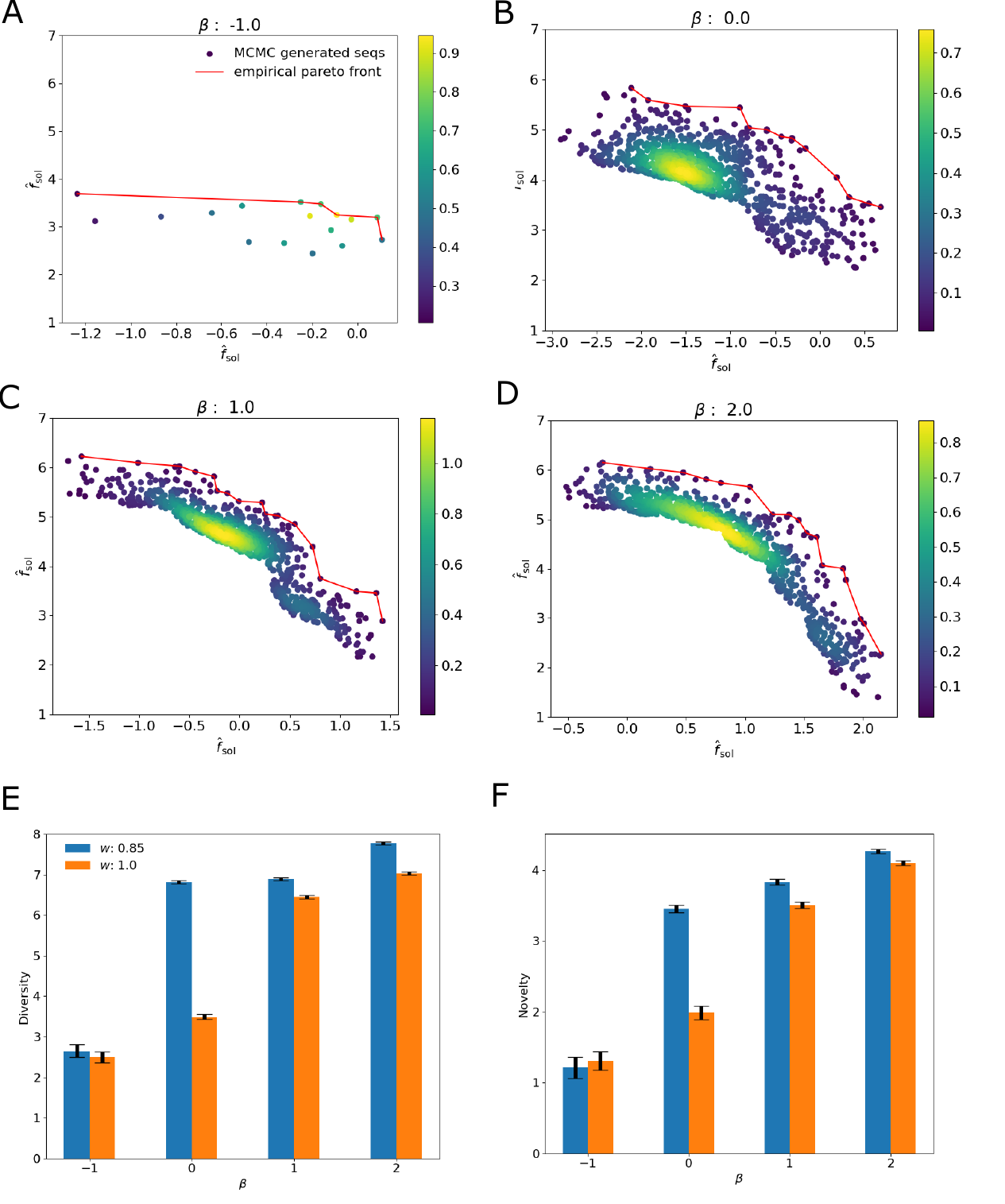}
\end{center}
\caption{A. Density plots of all distinct sequences sub-sampled from the set generated by Metropolis-Hastings at inverse temperature $T^{-1} = 10$ and $\beta = -1.0$ on the hard synthetic task. B. Density plots of 1000 sequences sub-sampled from the set generated by Metropolis-Hastings at inverse temperature $T^{-1} = 10$ and $\beta = 0.0$ on the simple synthetic task. C. Density plots of 1000 sequences sub-sampled from the set generated by Metropolis-Hastings at inverse temperature $T^{-1} = 10$ and $\beta = 1.0$ on the simple synthetic task. D. Density plots of 1000 sequences sub-sampled from the set generated by Metropolis-Hastings at inverse temperature $T^{-1} = 10$ and $\beta = 2.0$ on the simple synthetic task. E F. Comparison of diversity and novelty for different choices of $\beta$ and $w$.}
\label{figsi6}
\end{figure}

 % figsi7
\begin{figure}[H]
\begin{center}
  \includegraphics[width=\linewidth]{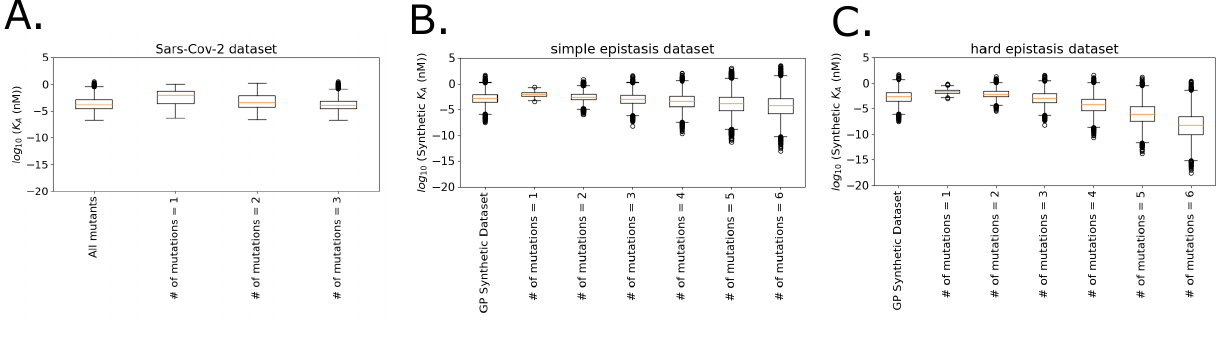}
\end{center}
\caption{A. Box plot showing the distribution of $K_A$ from the CB-119 binder dataset. B. Box plot showing the distribution of synthetic affinity $f_{\rm aff}$ for the dataset used to train the gaussian process. In addition, we generated 14000 random mutants from the wild type with 1 to 6 mutations and computed their synthetic affinity $f_{\rm aff}$ to estimate the distribution of $f_{\rm aff}$ conditioned on the number of mutations a mutant has. We show the box plots of the empirical distribution. C. Same as B. but for the hard synthetic task.  }
\label{figsi7}
\end{figure}

 % figsi9
\begin{figure*}
\begin{center}
\includegraphics[width=0.8\linewidth]{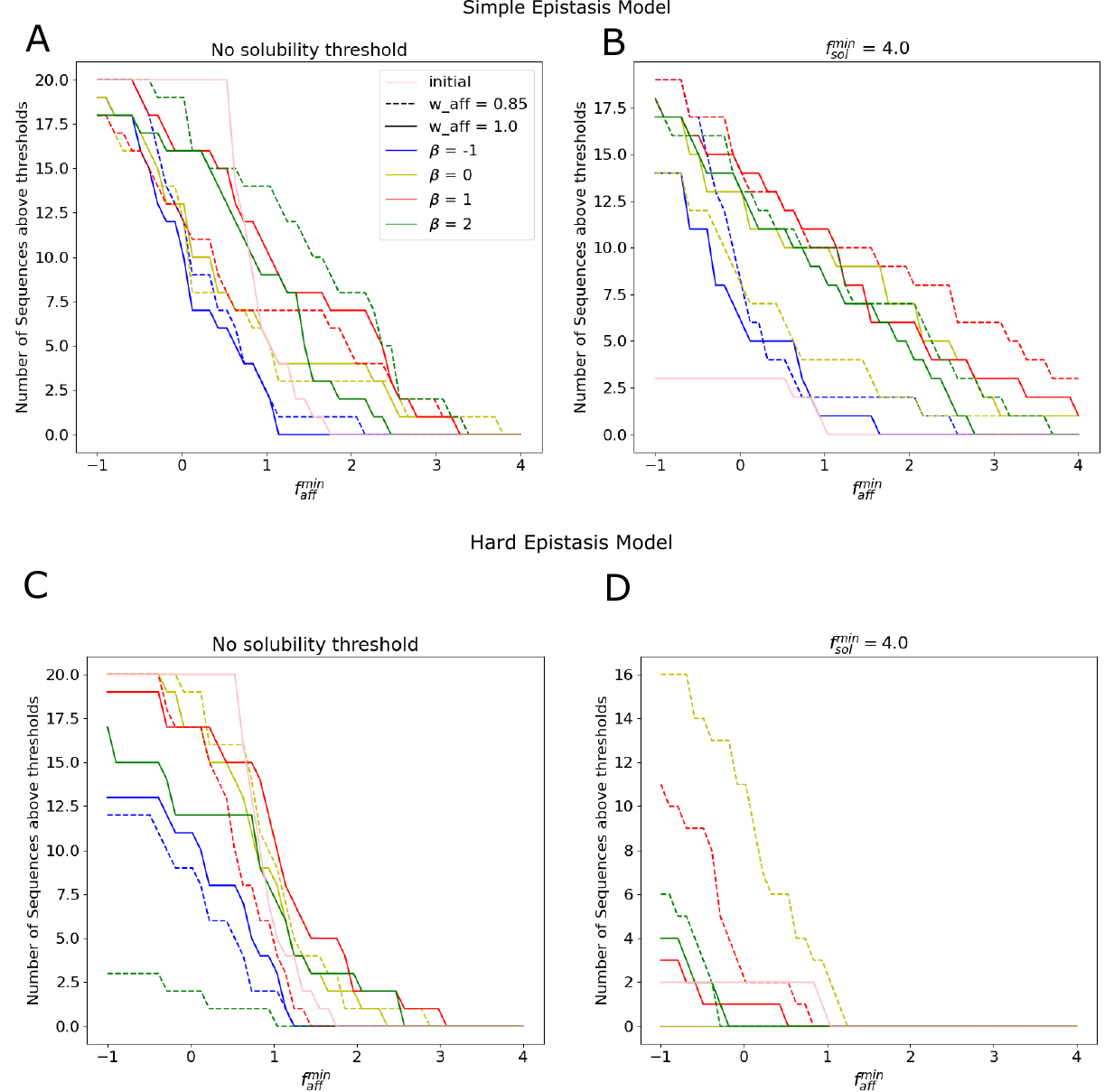}
\end{center}
\caption{
Figure A. shows the number of sequences out the top 20 selected from the set of generated sequences with Metropolis Hastings when trying to optimize the simple epistasis model with an affinity above a certain thresholds for different choices of $\beta$ and $w$. We compare those results to the set of sequences in the training set, indicated as initial. B shows the percentage of sequences that are above a certain threshold and also have a predicted solubility score above 3. Figure C. and D. show the results when trying to optimize the hard epistasis model.
}
\label{figsi:9}
\end{figure*}

 % figsi10
\begin{figure*}
\begin{center}
  \includegraphics[width=0.8\linewidth]{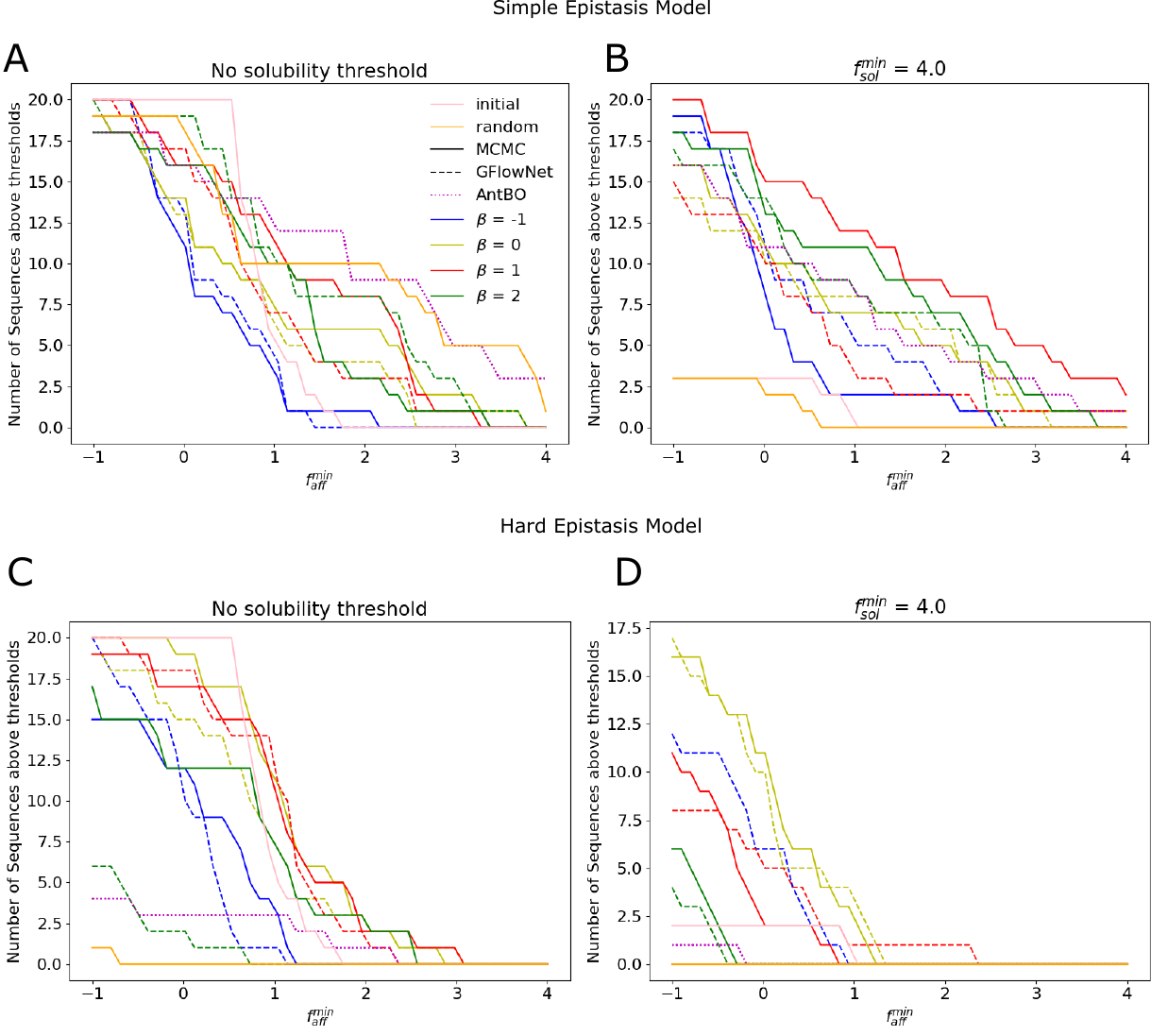}
\end{center}
\caption{
Figure A shows the number of sequences out the top 20 selected from the set of sequences generated for different choices of $\beta$ and generative method (Metropolis Hastings and GFlowNet). We compare those results to the set of sequences generated by the local search procedure of antBO and the set of sequences used to train the Gaussian Process affinity predictor, indicated as initial. B shows the percentage of sequences that are above a certain threshold and also have a predicted solubility score above 3. Figure C and D show the results when trying to optimize the hard epistasis model.
}
\label{figsi10}
\end{figure*}

\end{document}